\documentclass[prd,aps,superscriptaddress,preprintnumbers,notitlepage,nofootinbib]{revtex4-1}
\usepackage{amsmath,amssymb,graphicx,bm}
\usepackage{enumerate}
\usepackage{subfigure}
\usepackage{hyperref}

\newcommand{\beq}{\begin {equation}}  
\newcommand{\eeq}{\end   {equation}} 
\newcommand{\bea}{\begin {eqnarray}} 
\newcommand{\eea}{\end   {eqnarray}}  
\newcommand{\baa}{\begin {array}   } 
\newcommand{\eaa}{\end   {array}   }     
\newcommand{\bit}{\begin {itemize} }
\newcommand{\eit}{\end   {itemize} }
\newcommand{\be }{\begin {equation}} 
\newcommand{\ee }{\end   {equation}}

\usepackage{xcolor}

\definecolor{dgreen}{HTML}{008000}

\begin{document}

\preprint{ACFI-T17-10}

\title{Exploring Extended Scalar Sectors with di-Higgs Signals: A Higgs EFT Perspective}

\author{Tyler Corbett}
\affiliation{ARC Centre of Excellence for Particle Physics at the Terascale, School of Physics, The University of Melbourne, Victoria, Australia}
\author{Aniket Joglekar}
\affiliation{Enrico Fermi Institute and Kavli Institute for Cosmological Physics, University of Chicago, Chicago, IL 60637, USA}
\affiliation{Amherst Center for Fundamental Interactions, Department of Physics, University of Massachusetts Amherst, Amherst, MA 01003, USA} 
\author{Hao-Lin Li}
\affiliation{Amherst Center for Fundamental Interactions, Department of Physics, University of Massachusetts Amherst, Amherst, MA 01003, USA}
\author{Jiang-Hao Yu}
%\email{jhyu@itp.ac.cn}
\affiliation{CAS Key Laboratory of Theoretical Physics, Institute of Theoretical Physics, Chinese Academy of Sciences, Beijing 100190, P. R. China}
\affiliation{School of Physical Sciences, University of Chinese Academy of Sciences, No.19A Yuquan Road, Beijing 100049, P.R. China}
\affiliation{Amherst Center for Fundamental Interactions, Department of Physics, University of Massachusetts Amherst, Amherst, MA 01003, USA}

\begin{abstract}

We consider extended scalar sectors of the Standard Model as ultraviolet complete motivations for studying the effective Higgs self-interaction operators  of the Standard Model effective field theory. We investigate all motivated heavy scalar models which generate the dimension-six effective operator, $|H|^6$, at tree level and proceed to identify the full set of tree-level dimension-six operators by integrating out the heavy scalars. 
Of seven models which generate $|H|^6$ at tree level only two, quadruplets of hypercharge $Y=3Y_H$ and $Y=Y_H$, generate only this operator. 
Next we perform global fits to constrain relevant Wilson coefficients from the LHC single Higgs measurements as well as the electroweak oblique parameters $S$ and $T$. 
We find that the $T$ parameter puts very strong constraints on the Wilson coefficient of the $|H|^6$ operator in the triplet and quadruplet models, 
while the singlet and doublet models could still have Higgs self-couplings which deviate significantly from the standard model prediction.  
To determine the extent to which the $|H|^6$ operator could be constrained, we study the di-Higgs signatures at the future 100 TeV collider and explore future sensitivity of this operator. 
Projected onto the Higgs potential parameters of the extended scalar sectors, 
with $30$ ab$^{-1}$ luminosity data we will be able to explore the Higgs potential parameters in all seven models.

\end{abstract}

\maketitle

%%%%%%%%%%%%%%%%%%%%%%%%%%%%%%%%%%%%%%%%%%%%%%%%%%%%%%%%%%%%%%
\section{Introduction}
\label{sec:intro}
%%%%%%%%%%%%%%%%%%%%%%%%%%%%%%%%%%%%%%%%%%%%%%%%%%%%%%%%%%%%%%

The discovery of the Higgs boson at the Large Hadron Collider (LHC) marked the discovery of the last missing piece of the Standard Model (SM). 
Precision measurements of the Higgs couplings are a major goal for current and future high energy 
experiments. 
Current experimental results provide strong evidence that the nature of the Higgs boson is consistent with the predictions of the SM. 
The measurement of this behavior is entirely dependent on single Higgs phenomena through precision measurements of the Higgs couplings to the vector bosons and the SM fermions.
On the other hand, the Higgs self-interactions, responsible for Electroweak Symmetry Breaking (EWSB), still remain undetermined experimentally.
The Higgs self-coupling directly determines the shape of the Higgs potential and therefore measuring possible deviations of the Higgs self-coupling from its SM value 
is a crucial step in understanding the nature of EWSB, electroweak vacuum stability, and the nature of the electroweak phase transition (EWPT).

In order to investigate the generic features of the trihiggs coupling at the LHC and a future collider we adopt an Effective Field Theory (EFT) approach~\cite{Buchmuller:1985jz, Grzadkowski:2010es, Giudice:2007fh, Masso:2014xra}. 
In doing so we assume some possible new physics beyond the SM which modifies the Higgs couplings and is heavy with, for example, new physics scales such as $\Lambda_{\rm NP} \sim $ TeV. 
The effects of the new physics are parametrized by higher dimension effective operators, and 
the dimension-six $Q_H=(H^\dagger H)^3$ operator is the leading operator which modifies the momentum independent Higgs self-couplings at low energy.
The $Q_H$ operator remains the only operator related to the Higgs sector unconstrained by current experiment.
In order to motivate this study of the effective operator $Q_H$ we consider ultraviolet (UV) complete models which may generate this operator at tree level and therefore with a larger Wilson coefficient. 
This requirement combined with Lorentz invariance then limits our consideration to extended scalar sectors\footnote{Requiring closure of spinor and Lorentz indices implies fermions may only generate the $Q_H$ operator at one-loop and vectors may only generate dimension-six-operators with two derivatives at tree-level. }. 
%
%These scalars should carry some $SU(2)_L$ charge which allows for a tree-level $Q_H$ along with a corresponding hypercharge $Y$. 
%
Additionally the new scalar must not be charged under $SU(3)_c$ as closure of color indices requires $Q_H$ be generated at one loop. 
Such scalar extensions of the SM
constitute relatively simple scenarios beyond the SM which are also well-motivated by studies of the electroweak phase transition and baryogenesis \cite{Barger:2007im, Barger:2008jx}, 
having dark matter candidates \cite{McDonald:1993ex, Burgess:2000yq, FileviezPerez:2008bj}, or mechanisms for neutrino mass generation~\cite{Konetschny:1977bn, Magg:1980ut, Schechter:1980gr, Cheng:1980qt}. 
The complete list of the scalar extensions which generate a  tree-level $Q_H$ are real~\cite{McDonald:1993ex, Burgess:2000yq, OConnell:2006rsp, Barger:2007im} and complex singlets~\cite{Barger:2008jx}, the two Higgs doublet model~(2HDM)~\cite{Branco:2011iw,Lee:1973iz,Gunion:2002zf}, real~\cite{Blank:1997qa, FileviezPerez:2008bj} and complex~\cite{Konetschny:1977bn, Magg:1980ut, Schechter:1980gr, Cheng:1980qt} triplets, and complex quadruplets.  
Assuming the new scalars in these models are heavy, we utilize an EFT approach to study their effects on electroweak precision tests, modifications of the single Higgs couplings, and the di-Higgs production process in a model-independent and predictive way. 

Many new physics models with SM-compatible single Higgs phenomena could exhibit di-Higgs phenomenology distinct from that of the SM~\cite{deFlorian:2016spz, Azatov:2015oxa}.
The modifications of the Higgs trilinear couplings can only be directly observed in Higgs boson pair production, therefore the di-Higgs process at the LHC and future colliders is the only direct way to measure the Wilson coefficient of the effective $Q_H$ operator. 
Alternatively the trilinear Higgs coupling can be studied indirectly \cite{Degrassi:2016wml,DiVita:2017eyz,Maltoni:2017ims,Degrassi:2016wml,Gorbahn:2016uoy,Bizon:2016wgr,DiVita:2017eyz}. However our paper will focus on the direct constraints, we discuss the indirect constraints briefly at the end of Section~\ref{sec:model}.
The di-Higgs production mechanism at hadronic colliders is dominated by the  gluon fusion process which includes the triangle and the box contributions from the top quark. 
Due to destructive interference between these two contributions, the di-Higgs production cross section in the SM is typically small and thus challenging to observe in the near future. 
However, in the scalar extended models, the di-Higgs cross section may be increased considerably making measurement a possibility at the proposed 100 TeV collider~\cite{Arkani-Hamed:2015vfh, Contino:2016spe}.
In this paper we investigate the di-Higgs production cross sections in the EFT framework, and study the discovery potential of the Wilson coefficients in the EFT at the proposed 100 TeV collider.

We proceed by, in Section~\ref{sec:model}, listing the simplest scalar extensions of the SM which generate $Q_H$ at tree-level along with their Lagrangians and the corresponding effective Lagrangians after integrating out the new heavy degrees of freedom. Then in Section~\ref{sec:lhc} we study the implications of single-Higgs measurements on the corresponding EFTs as well as the implications of these constraints on the UV complete models. In Section~\ref{sec:dihiggs} we study these EFTs' impacts on di-Higgs production at the proposed 100 TeV collider. Finally our conclusions are found in Section~\ref{sec:conclusion}.

%%%%%%%%%%%%%%%%%%%%%%%%%%%%%%%%%%%%%%%%%%%%%%%%%%%%%%%%%%%%%%

\section{The Effective Lagrangian}\label{sec:model}
We consider all ultraviolet (UV) complete models which include one additional heavy scalar which generate, after integrating out the new scalar, dimension-six operators affecting the trihiggs vertex at tree level. In order to generate tree level dimension-six operators one needs a term $H^2 S$ or $H^3 S$, where S is the new heavy scalar, this is a result of all other models having an additional $\mathbb{Z}_2$ symmetry due to the requirements of the gauge symmetry and renormalizability\footnote{An exception to this is the $HS^3$ or $HS^2$ vertex, however the $HS^3$ vertex will not generate operators at tree level below dimension-eight and the $HS^2$ vertex does not exist for any representations given the Higgs is a doublet of $SU(2)_L$.}.

The relevant theories are then, real and complex scalar singlets, the two-Higgs doublet model, real and complex scalar triplets of $SU(2)_L$ with hypercharge $Y=0$ and $Y=-1$ respectively, and finally complex scalar quadruplets of $SU(2)_L$ with either $Y=3/2$ or $Y=1/2$. For each model we write down the Lagrangians for each UV-model along with the corresponding effective field theory (EFT) to dimension-six at tree level, we will only write the new terms in addition to the standard model terms for convenience. In writing the EFTs we will follow the procedure of Henning~et~al.~\cite{Henning:2014wua,Henning:2016lyp}. To clarify our notation and conventions, we write here the general Lagrangian for all UV complete models, neglecting SM fermionic and gauge boson terms, considered:
\begin{equation}\label{eq:uvabbreviated}
\mathcal{L}=(D^\mu H)^\dagger(D_\mu H)-\mu^2(H^\dagger H)-\lambda(H^\dagger H)^2+\Delta\mathcal{L}
\end{equation}
Where $\Delta\mathcal{L}$ contains all terms containing new fields (in the case of the models we consider this is one new scalar multiplet of $SU(2)$ which may or may not have hypercharge). $\mu^2$ becoming negative signals spontaneous symmetry breaking leading to the massive gauge bosons of the SM. After deriving the EFTs we employ the Warsaw basis~\cite{Grzadkowski:2010es} for the dimension-six operators, translations between the various bases are included throughout much of the recent literature including a package for relating the bases \cite{Falkowski:2015wza}. The operators which are relevant to our analyses are:
\begin{equation}
\begin{array}{rcl c rcl}
Q_H&=&(H^\dagger H)^3\, ,&\ \ \ \ \ &Q_{eH}&=&(H^\dagger H)(\bar L e_R H)\, ,\\
Q_{H\Box}&=&(H^\dagger H)\Box(H^\dagger H)\, ,&\ \ \ \ \ &Q_{uH}&=&(H^\dagger H)(\bar Q u_R \tilde H)\, ,\\
Q_{HD}&=&(D^\mu H)^\dagger HH^\dagger (D_\mu H)\, ,&\ \ \ \ \ &Q_{dH}&=&(H^\dagger H)(\bar Q d_R H)\, .
\end{array}
\end{equation}
The fermionic operators should be summed over each generation with an appropriate Wilson coefficient. In general the fermionic operators can have off diagonal components, however for the models considered this is only possible for the two-Higgs doublet model and we will employ particular choices of the fermionic matrices in the model to suppress off diagonal components, as is motivated by studies of flavor changing neutral currents, and therefore assume these operators to be diagonal.

\subsection{Real Scalar Singlet}
The real scalar singlet has $Y=0$, it has been studied extensively in the literature both from the UV complete~~\cite{McDonald:1993ex, Burgess:2000yq, OConnell:2006rsp, Barger:2007im} % 
and EFT perspectives \cite{Corbett:2015lfa,Henning:2014wua,Gorbahn:2015gxa}.
The Lagrangian, neglecting SM terms, is given by:
\begin{equation}\label{eq:dLscalarsinglet}
\Delta\mathcal{L}=\frac{1}{2}(\partial^\mu S)(\partial_\mu S)-\frac{M^2}{2}S^2-\frac{g}{3}S^3-g_{HS}(H^\dagger H)S-\frac{\lambda_S}{4}S^4-\frac{\lambda_{HS}}{2}(H^\dagger H)S^2\, .
\end{equation}
After integrating out the $S$ field we find the EFT:
\begin{equation}
\Delta\mathcal{L}\to\frac{g_{HS}^2}{2M^2}(H^\dagger H)^2-\left(\frac{\lambda_{HS}}{2}{-\frac{gg_{HS}}{3M^2}}\right)\frac{g_{HS}^2}{M^4}Q_H-\frac{g_{HS}^2}{2M^4}Q_{H\Box}\, .
\end{equation}
We note that there are corrections to the renormalizable $|H|^4$ vertex, which we will find is a common feature of integrating out scalars in our models, as well as the dimension-six operators $Q_H$ and $Q_{H\Box}$ which affect the trihiggs couplings. Additionally the term $g g_{HS}^3/3/M^6$ appears to be of the next order in the EFT expansion, we will retain these terms in the text, however in our summary Tables~\ref{tab:h4renorm}~and~\ref{tab:EFTtable} we neglect such corrections.

\subsection{Complex Scalar Singlet}
For the complex scalar singlet we consider the case of $Y=0$. While the complex scalar singlet is technically the same as introducing two real singlets, and therefore doesn't fit our criteria for considered models, we consider it here as it has been studied extensively in the literature. Some examples from the literature which study the complex singlet case and its implications for inflation, the electroweak phase transition, enhancement of the di-Higgs signal, and vacuum stability include~\cite{Barger:2008jx,Ferreira:2016tcu,Chiang:2017nmu,Dawson:2017jja,Cheng:2018ajh}. The Lagrangian is then:
\begin{eqnarray}
\Delta\mathcal{L}&=&(\partial^\mu\Phi)^\dagger(\partial_\mu\Phi)-M^2|\Phi|^2-\frac{(M')^2}{2}\left(\Phi^2+h.c.\right)\nonumber\\
&&-\left(g_{HS}(H^\dagger H)\Phi+h.c.\right)-\left(\frac{g}{3}\Phi^3+h.c.\right)-\left(\frac{g'}{3}\Phi(\Phi^\dagger)^2+h.c.\right)\nonumber\\
&& -\left(\frac{\lambda_{H\Phi}}{2}(H^\dagger H)\Phi^2+h.c.\right)-\frac{\lambda_{H\Phi}'}{2}(H^\dagger H) |\Phi|^2-\left(\frac{\lambda}{4}\Phi^4+h.c.\right)-\frac{\lambda'}{4}|\Phi|^4-\left(\frac{\lambda_1}{4}\Phi(\Phi^\dagger)^3+h.c.\right)
\end{eqnarray}
The $M'$ term corrects the dimension-six operator coefficients with terms proportional to $M'/M$ which must be small for the validity of the EFT so we neglect them\footnote{$M'$ is the parameter which dictates the size of the mass splitting between the components of the complex scalar field. If $M'$ were to become large it is possible that the lighter resonances would enter the low energy spectrum and invalidate our EFT approach. Therefore it is a requirement of our EFT approach that this parameter be small. For the same reason we will neglect the effects of $Y_3$ in the 2HDM below.}. Integrating out $\Phi$ and $\Phi^\dagger$ gives the effective Lagrangian:
\begin{equation}
\Delta\mathcal{L}\to\frac{|g_{HS}|^2}{M^2}(H^\dagger H)^2-\left(\frac{|g_{HS}|^2\lambda_{H\Phi}'}{2M^4}+\frac{{\rm Re}[g_{HS}^2\lambda_{H\Phi}]}{M^4}{-\frac{2{\rm Re}[g_{HS}^3g^*+g_{HS}^2g'g_{HS}]}{M^6}}\right)Q_H-\frac{|g_{HS}|^2}{M^4}Q_{H\Box}
\end{equation}
Again we induce corrections to the $|H|^4$ vertex as well as the effective operators $Q_H$ and $Q_{H\Box}$.
%

%
%SUBSECTION 2HDM
\subsection{Two Higgs Doublet Model}
\begin{table}
\begin{tabular}{|r|c|c|c|}
\hline
&L&U&D\\
\hline
Type I:	&$\Phi_2$	&$\Phi_2$	&$\Phi_2$	\\
Type II:	&$\Phi_1$	&$\Phi_2$	&$\Phi_1$	\\
Lepton-Specific:	&$\Phi_1$	&$\Phi_2$	&$\Phi_2$	\\
Flipped:	&$\Phi_2$	&$\Phi_2$	&$\Phi_1$	\\
\hline
\end{tabular}
\caption{List of Fermion couplings used for various Types of 2HDM.}\label{tab:2hdmtypes}
\end{table}
Of the many extended scalar sectors studied in the literature the two Higgs doublet model is the most well studied, reviews on the status of the model from the UV perspective have a long history (some extensive reviews include \cite{Branco:2011iw,Lee:1973iz,Gunion:2002zf}), the two Higgs doublet model has also recently been studied in the EFT framework in great detail \cite{Belusca-Maito:2016dqe,deBlas:2014mba,Gorbahn:2015gxa} including comparisons between the phenomenological aspects of both the UV complete and EFT frameworks at tree and one-loop levels \cite{Brehmer:2015rna,Freitas:2016iwx}.

We begin in the ``Higgs basis'', where the doublets have already been rotated to a basis where the physical CP even state is the observed 125 GeV Higgs. This rotation is performed by rotation of $H_1$ and $H_2$ by the angle $\beta$. We follow the notation of \cite{Belusca-Maito:2016dqe}. Note the Yukawa couplings are entered generically and later will be recast in terms of each of the four ``types'' usually considered to evade flavor changing neutral currents when we write the EFT. These various types considered are outlined in Table~\ref{tab:2hdmtypes}. 
\begin{eqnarray}\label{eq:dL2HDM}
\Delta\mathcal{L}&=&(D^\mu H_2)^\dagger(D_\mu H_2)-M^2|H_2|^2-Y_3(H_1^\dagger H_2+h.c.)-\frac{Z_2}{2}|H_2|^4-Z_3|H_1|^2|H_2|^2-Z_4(H_1^\dagger H_2)(H_2^\dagger H_1)\nonumber\\
&&-\frac{Z_5}{2}(H_1^\dagger H_2)(H_1^\dagger H_2)-\frac{Z_5^*}{2}(H_2^\dagger H_1)(H_2^\dagger H_1)-Z_6|H_1|^2(H_1^\dagger H_2)-Z_6^*|H_1|^2(H_2^\dagger H_1)-Z_7|H_2|^2(H_1^\dagger H_2)-Z_7^*|H_2|^2(H_2^\dagger H_1)\nonumber\\
&&-\left(H_{2,i} \bar Q_j Y_u u_R\epsilon_{ij}+H_{2,i} \bar Q_iY_d d_R+H_{2,i}\bar L_i Y_l e_R+h.c.\right)
\end{eqnarray}
The effective Lagrangian for each ``type'' of 2HDM is then given below. Note we have neglected terms suppressed by $Y_3/M^2$ as explained above in the complex scalar discussion. We adopt the notation $\cos\beta=c_\beta$ and $\sin\beta=s_\beta$, where the mixing angle $\beta$ is the angle which diagonalizes the mass matrices of the charged scalars and pseudoscalars, to allow us to rewrite the Higgs-fermion couplings in terms of the mixing angle and the parameter $Z_6$. 
\begin{itemize}
\item \textit{Type I:}
\begin{eqnarray}\label{eq:LT1}
\Delta\mathcal{L}
&=&\frac{Z_6}{M^2}\frac{2vh+h^2}{2}\left(\frac{\sqrt{2}m_lc_\beta}{vs_\beta}\bar LH_1 e_R+\frac{\sqrt{2}m_uc_\beta}{vs_\beta}\bar Q\tilde H_1 u_R+\frac{\sqrt{2}m_dc_\beta}{vs_\beta}\bar Q H_1 d_R+h.c.\right)\nonumber\\
&&+\frac{|Z_6|^2}{M^2}Q_H+\frac{1}{M^2}\left({\rm 4-Fermi}\right)
\end{eqnarray}
\item \textit{Type II:}
\begin{eqnarray}\label{eq:LT2}
\Delta\mathcal{L}
&=&\frac{Z_6}{M^2} \frac{2vh+h^2}{2}\left(-\frac{\sqrt{2} m_ls_\beta}{vc_\beta} \bar LH_1 e_r+\frac{\sqrt{2} m_uc_\beta}{vs_\beta} \bar Q\tilde H_1 u_R-\frac{\sqrt{2} m_ds_\beta}{vc_\beta} \bar Q H_1 d_R+h.c.\right)\nonumber\\
&&+\frac{|Z_6|^2}{M^2}Q_H+\frac{1}{M^2}\left({\rm 4-Fermi}\right)
\end{eqnarray}
\item \textit{Lepton Specific:}
\begin{eqnarray}\label{eq:LTLeptonSpecific}
\Delta\mathcal{L}
&=&\frac{Z_6}{M^2}\frac{2vh+h^2}{2}\left(-\frac{\sqrt{2} m_ls_\beta}{vc_\beta} \bar LH_1 e_r+\frac{\sqrt{2} m_uc_\beta}{vs_\beta} \bar Q\tilde H_1 u_R+\frac{\sqrt{2}m_dc_\beta}{vs_\beta}\bar Q H_1 d_R+h.c.\right)\nonumber\\
&&+\frac{|Z_6|^2}{M^2}Q_H+\frac{1}{M^2}\left({\rm 4-Fermi}\right)
\end{eqnarray}
\item \textit{Flipped:}
\begin{eqnarray}\label{eq:LTFlipped}
\Delta\mathcal{L}
&=&\frac{Z_6}{M^2}\frac{2vh+h^2}{2}\left(\frac{\sqrt{2} m_lc_\beta}{vs_\beta} \bar LH_1 e_r+\frac{\sqrt{2} m_uc_\beta}{vs_\beta} \bar Q\tilde H_1 u_R-\frac{\sqrt{2}m_ds_\beta}{vc_\beta} \bar Q H_1 d_R+h.c.\right)\nonumber\\
&&+\frac{|Z_6|^2}{M^2}Q_H+\frac{1}{M^2}\left({\rm 4-Fermi}\right)
\end{eqnarray}
\end{itemize}
We see that the 2HDM only induces one purely bosonic operator, $Q_H$, at leading order in $Y_3/M^2$, and induces various combinations of rescalings of the Yukawa couplings, i.e. the operators $Q_{eH}$, $Q_{uH}$, and $Q_{dH}$. The only difference between the various realizations of the 2HDM considered are differences in the weight of the fermionic operators, i.e. by $\tan\beta$ or $\cot\beta$. To make manifest the mass dependence of the Higgs couplings to fermions above we have expanded the fermionic dimension-six operators (in the unitary gauge for convenience) to recast the couplings of $H_1$ to fermions in terms of their masses, $Z_6$, and the mixing angle $\beta$. In particular the first line of each expression indicates the shift of the Higgs-fermion couplings relative to the SM prediction,
\begin{equation}
\mathcal{L}_{H\psi\psi}=\frac{\sqrt{2}m_\psi}{v}h\bar\psi_R\psi_L\, .
\end{equation}
Another unique feature of the 2HDM effective Lagrangians is that they also contain 4-Fermi operators. These are not relevant to our analysis and, as they are weighted by the square of the Yukawa, are unlikely to have large Wilson coefficients except possibly in the case of the top quark which has $Y_t\sim1$.

\subsection{Real Scalar Triplet}
The real scalar triplet model~\cite{Blank:1997qa, Chen:2006pb, SekharChivukula:2007gi} has been studied in the literature with ambitions of making the electroweak phase transition first order, e.g. in \cite{Inoue:2015pza}, with 
the possibility of the neutral component being a dark matter candidate~\cite{FileviezPerez:2008bj}, as well as from an EFT point of view in \cite{Khandker:2012zu, Henning:2014wua}.

The relevant Lagrangian is given by,
\begin{equation}
\Delta\mathcal{L}=\frac{1}{2}(D_\mu\Phi^a)^2-\frac{1}{2}M^2\Phi^a\Phi^a+gH^\dagger\tau^a H\Phi^a-\frac{\lambda_{H\Phi}}{2}(H^\dagger H)\Phi^a\Phi^a-\frac{1}{4}\lambda_\Phi(\Phi^a\Phi^a)^2\, .
\end{equation}
Integrating out the heavy triplet then gives the effective Lagrangian:
\begin{equation}\label{eq:RSTripletEFT1}
\Delta\mathcal{L}
=\frac{g^2}{8M^2}(H^\dagger H)^2-\frac{g^2}{2M^4}Q_{HD}-\frac{g^2}{8M^4}Q_{H\Box}+\frac{g^2}{2M^4}(H^\dagger H)(D_\mu H)^\dagger(D^\mu H)-\frac{g^2\lambda_{H\Phi}}{8M^4}Q_H\, .
\end{equation}
It is convenient to make a change of basis here, we may exchange the operator $|H|^2(D_\mu H)^\dagger(D^\mu H)$ for the other dimension-six operators at the cost of an error of the next order in the EFT (i.e. $\mathcal{O}(1/\Lambda^4)$). While it is frequently simpler to maintain the basis obtained after integrating out the heavy states \cite{Wudka:1994ny}, for the sake of this work which will consider many UV completions and their effective field theories we choose to project onto a common basis. Discussions of the validity of this method including proofs of the invariance of the S-matrix can be found in \cite{Politzer:1980me,Georgi:1991ch,Arzt:1993gz,Simma:1993ky}. We perform the change of basis by using the Higgs equation of motion, scaled up to dimension-six through multiplication by additional Higgs fields,
\begin{equation}\label{eq:EOMH}
(H^\dagger H)(D^\mu H)^\dagger(D_\mu H)=-{\lambda_R v^2}(H^\dagger H)^2+\frac{1}{2}Q_{H\Box}+2\lambda_R Q_H+\frac{1}{2}(Y_lQ_{lH}+Y_dQ_{dH}+Y_uQ_{uH}+h.c.)+\mathcal{O}(1/\Lambda^4)\, ,
\end{equation}
where we have called the renormalized $(H^\dagger H)^2$ coupling, $\lambda_R=\lambda+g^2/8/M^2$ with $\lambda$ the $(H^\dagger H)^2$ coupling of Eq.~\ref{eq:uvabbreviated}, yielding the new form of Eq.~\ref{eq:RSTripletEFT1}:
\begin{eqnarray}
\Delta\mathcal{L}&=&\frac{g^2}{M^2}\left(\frac{1}{8}-\frac{\lambda v^2}{2M^2}{-\frac{g^2v^2}{16M^4}}\right)(H^\dagger H)^2-\frac{g^2}{2M^4}Q_{HD}+\frac{g^2}{8M^4}Q_{H\Box}-\frac{g^2}{M^4}\left(\frac{\lambda_{H\Phi}}{8}-\lambda{-\frac{g^2}{8M^2}}\right)Q_H\nonumber\\
&&+\frac{g^2}{4M^4}\left(Y_lQ_{lH}+Y_dQ_{dH}+Y_uQ_{uH}+h.c.\right)\, .\label{eq:EFTRtripletfinal}
\end{eqnarray}
Consistent with our other examples we have again generated the $Q_H$ and $Q_{H\Box}$ operators, however interestingly we have also generated the $Q_{HD}$ operator which will have important phenomenological implications which we discuss in Section~\ref{sec:lhc}.

\subsection{Complex Scalar Triplet}
Charging the Scalar Triplet under hypercharge, $Y=-1$, has important uses in the Type II seesaw~\cite{Konetschny:1977bn, Magg:1980ut, Schechter:1980gr, Cheng:1980qt}. 
  The relevant UV complete Lagrangian is then,
\begin{eqnarray}\label{eq:dLCscalarTrip}
\Delta\mathcal{L}&=&|D_\mu\Phi^a|^2-M^2|\Phi^a|^2+(g H^Ti\sigma_2\tau^a H\Phi^a+h.c.)\nonumber\\
&&-\frac{\lambda_{H\Phi}}{2}|H|^2|\Phi^a|^2-\frac{\lambda'}{4}(H^\dagger\tau^a\tau^b H)\Phi^a(\Phi^b)^\dagger-\frac{1}{4}\lambda_\Phi|\Phi^a|^4-\frac{1}{4}\lambda_{\Phi}'{\rm Tr}[\tau^a\tau^b\tau^c\tau^d](\Phi^a)^\dagger\Phi^b(\Phi^c)^\dagger\Phi^d\, .
\end{eqnarray}
Integrating out the heavy complex triplet yields the effective Lagrangian,
\begin{equation}
\Delta\mathcal{L}=\frac{|g|^2}{2M^2}(H^\dagger H)^2+\frac{|g|^2}{M^4}(H^\dagger H)(D^\mu H)^\dagger(D_\mu H)+\frac{|g|^2}{M^4} Q_{HD}
-\frac{|g|^2}{2M^4}\left(\frac{\lambda_{H\Phi}}{2}+\frac{\lambda'}{4}\right)Q_H\, ,
\end{equation}
which after applying the equation of motion from Eq.~\ref{eq:EOMH} (notice here $\lambda_R=\lambda+|g^2|/2/M^2$) gives the final form for the effective Lagrangian:
\begin{eqnarray}
\Delta\mathcal{L}&=&\frac{|g|^2}{M^2}\left(\frac{1}{2}-\frac{\lambda v^2}{M^2}{-\frac{|g|^2v^2}{2M^4}}\right)(H^\dagger H)^2+\frac{|g|^2}{2M^4}Q_{H\Box}+\frac{|g|^2}{M^4}Q_{HD}-\frac{|g|^2}{M^4}\left(\frac{\lambda_{H\Phi}}{4}+\frac{\lambda'}{8}-2\lambda{-\frac{|g|^2}{M^2}}\right)Q_H\nonumber\\
&&+\frac{|g|^2}{2M^4}\left(Y_lQ_{lH}+Y_dQ_{dH}+Y_uQ_{uH}+h.c.\right)\, .\label{eq:EFTCtripletfinal}
\end{eqnarray}
This effective Lagrangian and the effective operators it contains are consistent with our expectations from the other models, particularly the real scalar triplet.

\subsection{Quadruplet with $Y=3Y_H$}
For the two quadruplet models we follow the notation of \cite{Hisano:2013sn}, the UV Lagrangian is then given by:
\begin{eqnarray}
\Delta \mathcal{L}&=&(D^\mu \Phi^{*ijk})(D_\mu \Phi_{ijk})-M^2\Phi^{*ijk}\Phi_{ijk}-(\lambda_{H3\Phi }H^{*i}H^{*j}H^{*k}\Phi_{ijk}+h.c.)\nonumber\\
&&-\lambda_{H2\Phi2}H^{*i}H_i\Phi^{*lmn}\Phi_{lmn}-\lambda'_{H2\Phi2}H^{*i}\Phi_{ijk}\Phi^{*jkl}H_l-\lambda_\Phi(\Phi^{*ijk}\Phi_{ijk})^2-\lambda_\Phi'(\Phi^{*ijk}\Phi_{ilm}\Phi^{*lmn}\Phi_{jkn})\, .
\end{eqnarray}
Integrating out the quadruplet leads to the simple EFT,
\begin{eqnarray}
\Delta\mathcal{L}&=&\frac{|\lambda_{H3\Phi}|^2}{M^2}(H^\dagger H)^3\, .
\end{eqnarray}
Note that for a quadruplet we expect a contribution to the $T$-parameter. This operator does not occur at dimension-six, but does at dimension-eight. Deriving only the dimension-eight operator contributing to the $T$-parameter yields:
\begin{equation}\label{eq:d8T3}
\mathcal{L}_8^T=\frac{6|\lambda_{H3\Phi}|^2}{M^4}|H^\dagger D^\mu H|^2|H|^2
\end{equation}
Here we have confirmed the sign of \cite{Dawson:2017vgm}. We will see in the case of $Y=Y_H$ we obtain a different sign from this work.

\subsection{Quadruplet with $Y=Y_H$}
The UV complete Lagrangian is given by,
\begin{eqnarray}
\Delta \mathcal{L}&=&(D^\mu \Phi^{*ijk})(D_\mu \Phi_{ijk})-M^2\Phi^{*ijk}\Phi_{ijk}-(\lambda_{H3\Phi }H^{*i}\Phi_{ijk}H^{*j}\epsilon^{kl}H_l+h.c.)\nonumber\\
&&-\lambda_{H2\Phi2}H^{*i}H_i\Phi^{*lmn}\Phi_{lmn}-\lambda'_{H2\Phi2}H^{*i}\Phi_{ijk}\Phi^{*jkl}H_l-\lambda_\Phi(\Phi^{*ijk}\Phi_{ijk})^2-\lambda_\Phi'(\Phi^{*ijk}\Phi_{ilm}\Phi^{*lmn}\Phi_{jkn})\, .
\end{eqnarray}
Again we find a very simple EFT to dimension-six:
\begin{eqnarray}\label{eq:d8T1}
\Delta\mathcal{L}&=&\frac{|\lambda_{H3\Phi}|^2}{M^2}(H^\dagger H)^3\, .
\end{eqnarray}
which we supplement with the dimension-eight $T$-parameter operator.
\begin{equation}
\mathcal{L}_8^T=\frac{2|\lambda_{H3\Phi}|^2}{M^4}|H^\dagger D^\mu H|^2|H|^2
\end{equation}
This expression agrees with \cite{Dawson:2017vgm} up to a sign. As the sign of the dimension-eight $T$ parameter operators in each quadruplet model come purely from the covariant derivative term of the Lagrangians (other contributions cancel) they should be the same in both Eqs.~\ref{eq:d8T3}~and~\ref{eq:d8T1}.

\subsection{Summary of EFTs}
Finally after deriving the corresponding EFTs for each model we may construct a table with the Wilson coefficients for each operator for each model considered. We summarize the renormalization of the $(H^\dagger H)^2$ term in Table~\ref{tab:h4renorm} and the Wilson coefficients of the dimension-six operators in Table~\ref{tab:EFTtable}. While it appears that of all the theories the 2HDM is the only which does not generate a correction to the renormalizable $(H^\dagger H)^2$, this is a reflection of neglecting terms suppressed by $Y_3/M^2$, these corrections are generated first at $\mathcal{O}(Y_3/M^2)$. Unsurprisingly neither the 2HDM nor the two singlet models generate $Q_{HD}$, also referred to as the $T$-parameter operator as they are known not to shift the relation between the $W$- and $Z$-masses. It is, however, expected from studies of the dynamics of the triplet models below EWSB that the triplet models considered in this work correct the $T$-parameter. This is consistent with our findings in Equations~\ref{eq:EFTRtripletfinal}~and~\ref{eq:EFTCtripletfinal}.
In the case of the quadruplet we found they were unique in that at dimension-six they generate only one operator, $Q_H$, and that the $T$-parameter operator was generated at dimension-eight. Additionally, as there are no allowed tree level couplings to Fermions in any of the theories except the 2HDM none of the other theories generate the fermionic operators, however after trading the operator $(H^\dagger H)(D^\mu H)^\dagger(D_\mu H)$ in the triplet models via the EOM we do generate the fermionic operators for the two triplet models.

The case of the quadruplets is particularly interesting as studies which indirectly probe the Higgs self coupling, such as \cite{Degrassi:2016wml}, only allow the SM coupling $\lambda$ to vary. Our work indicates that, within the assumptions of our EFT%
\footnote{For example relaxing the assumptions of a single new multiplet one could envision a scenario with multiple quadruplets in which the $T$-parameter bounds may be evaded allowing for a sizable $H^6$ operator coefficient and no other operators at dimension-six. In the case where only the $H^6$ operator is generated the indirect constraints may be more stringent than those of di-Higgs production \cite{Gorbahn:2016uoy}.}, 
such a study corresponds to a very specific UV complete scenario, in the case where one expects the NP to come from dimension-six operators this corresponds to the quadruplets. In the case of the quadruplets the shift in $\lambda$ due to the effective operators is restricted to be extremely small since the same UV parameter that generates the operator $Q_H$ contributes to the strongly constrained $T$-parameter. This demonstrates that indirect probes of the Higgs self coupling which don't vary other Higgs couplings are incomplete or correspond to specific UV completions which do not satisfy the criterion of the UV complete models considered. Other studies which vary these additional couplings of the Higgs such as \cite{DiVita:2017eyz,Maltoni:2017ims} indicate the bounds on the Higgs self coupling are weakened or even lost without the inclusion of the direct di-Higgs probe. 
\begin{table}
\bgroup
\def\arraystretch{1.8}
\begin{tabular}{| c|| c|}
\hline
\textbf{Theory:}&$\lambda_{RF}=\lambda+\cdots$\\
\hline
\hline
$\mathbb{R}$ Singlet&$\frac{g_{HS}^2}{2M^2}$\\
\hline
$\mathbb{C}$ Singlet&$\frac{|g_{HS}|^2}{M^2}$\\
\hline
2HDM&$0$\\
\hline
$\mathbb{R}$ Triplet ($Y=0$)&$\frac{g^2}{M^2}\left(\frac{1}{8}-\frac{\lambda v^2}{2M^2}\right)$\\
\hline
$\mathbb{C}$ Triplet ($Y=-1$)&$\frac{|g|^2}{M^2}\left(\frac{1}{2}-\frac{\lambda v^2}{M^2}\right)$\\
\hline
$\mathbb{C}$ Quadruplet ($Y=1/2$)&0\\
\hline
$\mathbb{C}$ Quadruplet ($Y=3/2$)&0\\
\hline
\end{tabular}
\egroup
\caption{Summary of the tree-level renormalization of the $(H^\dagger H)^2$ operator in the effective field theory. $\lambda_{RF}$ indicates the final renormalized $(H^\dagger H)^2$ coupling (i.e. after shifting the operators by the EOM) including $\lambda$ from Eq.~\ref{eq:uvabbreviated}. In this Table, as mentioned in the text in the Real Scalar singlet discussion, we neglect terms which are of $\mathcal{O}(g^4/M^6)$.}\label{tab:h4renorm}
\end{table}
\begin{table}
\bgroup
\def\arraystretch{1.8}
\begin{tabular}{| c || c | c | c | c | c | c |}
\hline
\textbf{Theory:}&$\mathbf{c_H}$&$\mathbf{c_{H\Box}}$&$\mathbf{c_{HD}}$&$\mathbf{c_{eH}}$&$\mathbf{c_{uH}}$&$\mathbf{c_{dH}}$\\
\hline
\hline
$\mathbb{R}$ Singlet&$-\frac{\lambda_{HS}}{2}\frac{g_{HS}^2}{M^4}$ 
&$-\frac{g_{HS}^2}{2M^4}$&-&-&-&-\\
\hline
$\mathbb{C}$ Singlet&$-\left(\frac{|g_{HS}|^2\lambda'_{H\Phi}}{2M^4}+\frac{{\rm Re}[g_{HS}^2\lambda_{H\Phi}]}{M^4}\right)$&$-\frac{|g_{HS}|^2}{M^4}$&-&-&-&-\\
\hline
2HDM, \hfill Type I&$\frac{|Z_6|^2}{M^2}$&-&-&$\frac{Z_6}{M^2}Y_lc_\beta$&$\frac{Z_6}{M^2}Y_uc_\beta$&$\frac{Z_6}{M^2}Y_dc_\beta$\\
\hfill Type II:&$\frac{|Z_6|^2}{M^2}$&-&-&$-\frac{Z_6}{M^2}Y_ls_\beta$&$\frac{Z_6}{M^2}Y_uc_\beta$&$-\frac{Z_6}{M^2}Y_ds_\beta$\\
\hfill Lepton-Specific:&$\frac{|Z_6|^2}{M^2}$&-&-&$-\frac{Z_6}{M^2}Y_ls_\beta$&$\frac{Z_6}{M^2}Y_uc_\beta$&$\frac{Z_6}{M^2}Y_dc_\beta$\\
\hfill Flipped:&$\frac{|Z_6|^2}{M^2}$&-&-&$\frac{Z_6}{M^2}Y_lc_\beta$&$\frac{Z_6}{M^2}Y_uc_\beta$&$-\frac{Z_6}{M^2}Y_ds_\beta$\\
\hline
$\mathbb{R}$ Triplet ($Y=0$)&$-\frac{g^2}{M^4}\left(\frac{\lambda_{H\Phi}}{8}-\lambda\right)$&$\frac{g^2}{8M^4}$&$-\frac{g^2}{2M^4}$&$\frac{g^2}{4M^4}Y_l$&$\frac{g^2}{4M^4}Y_u$&$\frac{g^2}{4M^4}Y_d$\\
\hline
$\mathbb{C}$ Triplet ($Y=-1$)&$-\frac{|g|^2}{M^4}\left(\frac{\lambda_{H\Phi}}{4}+\frac{\lambda'}{8}-2\lambda\right)$&$\frac{|g|^2}{2M^4}$&$\frac{|g|^2}{M^4}$&$\frac{|g|^2}{2M^4}Y_l$&$\frac{|g|^2}{2M^4}Y_u$&$\frac{|g|^2}{2M^4}Y_d$\\
\hline
$\mathbb{C}$ Quadruplet ($Y=1/2$)&$\frac{|\lambda_{H3\Phi}|^2}{M^2}$&-&$\frac{2|\lambda_{H3\Phi}|^2v^2}{2M^4}$&-&-&-\\
\hline
$\mathbb{C}$ Quadruplet ($Y=3/2$)&$\frac{|\lambda_{H3\Phi}|^2}{M^2}$&-&$\frac{6|\lambda_{H3\Phi}|^2v^2}{2M^4}$&-&-&-\\
\hline
\end{tabular}
\egroup
\caption{Summary of the tree-level effective field theory to dimension-six for the scalar theories considered. ``-'' indicates the operator is not generated in this theory. The UV operators with normalizations corresponding to each coupling constant should be read directly from the relevant Lagrangians in text. In this Table, as mentioned in the text in the Real Scalar singlet discussion, we neglect terms which are of $\mathcal{O}(g^4/M^6)$. While the operator $Q_{HD}$ is not generated in the quadruplet models we have entered the contributions to the $T$ parameter in terms of an effective coefficient for this operator into the table.}\label{tab:EFTtable}
\end{table}

It is useful to project these effective Lagrangians into Lorentz forms relevant to the di-Higgs analysis performed. We do so here, from the perspective of arbitrary Wilson coefficients, when the final analyses are performed we use the expressions for the Wilson coefficients expressed in Table~\ref{tab:EFTtable}. We assume that only the heaviest generation for each fermion has a non-negligible contribution to the EFT. Starting from the effective Lagrangian,
\begin{equation}\label{eq:fulleffectivelagrangian}
\mathcal{L}=(D^\mu H)^\dagger(D_\mu H)+|\mu|^2(H^\dagger H)-\lambda_{RF}(H^\dagger H)^2+c_HQ_H+c_{H\Box}Q_{H\Box}+c_{HD}Q_{HD}+c_{eH}Q_{eH}+c_{uH}Q_{uH}+c_{dH}Q_{dH}\, ,
\end{equation}
we can proceed to expand the operators to find the relevant Lorentz forms. Here we have used $\lambda_{RF}$ to represent the final renormalized coefficient of the $(H^\dagger H)^2$ operator, the expression for $\lambda_{RF}$ may be found in Table~\ref{tab:h4renorm} in terms of $\lambda$ of Eq.~\ref{eq:uvabbreviated} and the parameters of each UV-model. This involves finite field renormalizations as the operators $Q_{H\Box}$ and $Q_{HD}$ both alter the Higgs kinetic term below EWSB. Details of this procedure may be found in, for example,~\cite{Corbett:2012ja,Corbett:2014ora,Alonso:2013hga}. Below EWSB expanding out the Lorentz forms we find (employing the unitary gauge):
\begin{eqnarray}
\mathcal{L}&=&g_{HZZ}^{(3)}hZ_\mu Z^\mu
+g_{HWW}h W_\mu^+W^{-\mu}
+g_{HHH}^{(1)}h^3+g_{HHH}^{(2)}h(\partial_\mu h)(\partial^\mu h)\nonumber\\
\nonumber\\
&&+\left(g_{He}h\bar e_Le_R+g_{Hu}h \bar u_Lu_R+g_{Hd}h \bar d_L d_R+h.c.\right)
+\left(g_{HHu}h^2\bar u_Lu_R+h.c.\right)
+\cdots\, .\label{eq:lorentzlagrangian}
\end{eqnarray}
Here ``$\cdots$'' indicates the various operators and Lorentz forms which have no impact on our analysis. The coefficients of the terms in the Lagrangian of Eq.~\ref{eq:lorentzlagrangian} are given by: 
\begin{eqnarray}
g_{HWW}&=&{2}m_W^2(\sqrt{2}G_F)^{1/2}\left[1-\frac{v^2}{4}(c_{HD}-4c_{H\Box})\right]\nonumber\\
g_{HZZ}&=&m_Z^2(\sqrt{2}G_F)^{1/2}\left[1+\frac{v^2}{4}(c_{HD}+4c_{H\Box})\right]\nonumber\\
g_{HHH}^{(1)}&=&-\frac{m_H^2}{2}(\sqrt{2}G_F)^{1/2}\left[1-\frac{v^2}{4}(c_{HD}-4c_{H\Box}+{\frac{4}{\lambda_{RF}}c_H})\right]		\label{eq:lorentzcoeff}\\
g_{HHH}^{(2)}&=&\frac{1}{2(\sqrt{2}G_F)^{1/2}}(c_{HD}-4c_{H\Box})		\nonumber\\
g_{H\psi}&=&-m_\psi(\sqrt{2}G_F)^{1/2}\left[1-\frac{v^2}{4}(c_{HD}-4c_{H\Box})\right]+\frac{c_{\psi H}v^2}{\sqrt{2}}\, ,\nonumber\\
g_{HHu}&=&\frac{3c_{u H}}{2}\frac{v}{\sqrt{2}}\, .\nonumber
\end{eqnarray}
Note in Eq.~\ref{eq:lorentzcoeff} we have introduced $m_\psi$ and $c_{\psi H}$ as placeholders for the relevant fermion type (i.e. $e$, $u$, or $d$), and in this analysis we only consider couplings to the third generation of each. We have only included $g_{HHu}$ and its corresponding operator as only the top quark $h^2\bar\psi\psi$ operator will have an effect on our analyses as it is proportional to the top-quark Yukawa coupling which is the only large Yukawa in the SM. It is possible to remove the  $g_{HHH}^{(2)}$ operator by a field redefinition of $h$, however as pointed out in \cite{Passarino:2016saj} removing this operator by a field redefinition of $h$ (not the full doublet $H$) requires a nonlinear field redefinition which may prove to make one loop calculations difficult and if done incorrectly gauge dependent. Therefore we retain the $g_{HHH}^{(2)}$ coupling in favor of easier comparison with other works, such as those which study globally the constraints on the $h^3$ coupling via one loop dependent processes \cite{Kribs:2017znd,Degrassi:2017ucl,Degrassi:2016wml,Gorbahn:2016uoy,Bizon:2016wgr,DiVita:2017eyz}.

%%%%%%%%%%%%%%%%%%%%%%%%%%%%%%%%%%%%%%%%%%%%%%%%%%%%%%%%%%%%%%
\section{Higgs Coupling Measurements at the LHC}\label{sec:lhc}
In this section we consider important constraints on our EFTs in Section~\ref{sec:model}. We begin by considering the constraints from electroweak precision data along with a discussion of the loop order at which the $S$- and $T$-operators are generated either explicitly via integrating out at the mass scale of the extended scalar sectors or via operator mixing in the EFT while running down to the Higgs mass scale. Next we introduce the effective $h\gamma\gamma$ coupling in order to add an additional constraint to our global fit to single Higgs processes. We delegate to Appendix~\ref{sec:unitarity} unitarity considerations from the EFT perspective, where many amplitudes grow with the square of the center-of-mass energy $S$, as they do not add additional constraints to our models. Finally with our precision constraints on the EFTs we project these constraints into the UV complete models parameter spaces, this is especially useful in helping to limit the size of the $c_H$ coupling which is partially dependent on the same couplings as the $h\gamma\gamma$ effective coupling.

%%%%
\subsection{Electroweak Precision Measurements}\label{subsec:ewpmeasurements}

Electroweak precision data (EWPD) provide very strong constraints on the Wilson coefficients of effective operators. 
We note that 
the operator $Q_{HD}$ contributes at tree level to the $T$-parameter, while the operator,
\begin{equation}
Q_{HWB}=H^\dagger B^{\mu\nu}W_{\mu\nu}H\, ,
\end{equation}
contributes to the $S$-parameter at tree level. 
However, the only operators contributing to EWPD that are generated at tree- or one-loop level in our theories are $Q_{H\Box}$ and $Q_{HD}$ the operator $Q_{HWB}$ is only generated at two-loop or higher order. 
From Jenkins et al.~\cite{Jenkins:2013wua,Jenkins:2013zja,Alonso:2013hga} we have the elements of the anomalous dimension matrix for each of these operators:
\begin{eqnarray}
\dot c_H&=&\left(-\frac{27}{2}g_2^2-\frac{9}{2}g_1^2\right)c_H+\lambda\left[\frac{40}{3}g_2^2c_{H\Box}+(-6g_2^2+24 g_1^2 y_h^2)c_{HD}\right]+\cdots\nonumber\\
\dot c_{H\Box}&=&-\left(4g_2^2+\frac{16}{3}g_1^2y_h^2\right)c_{H\Box}+\frac{20}{3}g_1^2y_h^2c_{HD}+\cdots\label{eq:anomdimmatrix}\\
\dot c_{HD}&=&\frac{80}{3}g_1^2y_h^2 c_{H\Box}+\left(\frac{9}{2}g_2^2-\frac{10}{3}g_1^2y_h^2\right)c_{HD}+\cdots\nonumber\\
\dot c_{HWB}&=&6g_1g_2^2c_W+\left[-2y_h^2g_1^2+\frac{9}{2}g_2^2-\left(-\frac{1}{6}-\frac{20}{9}n_g\right)g_1^2-\left(\frac{43}{6}-\frac{4}{3}n_g\right)g_2^2\right]c_{HWB}+4g_1g_2y_hc_{HB}+4g_1g_2y_hc_{HW}\nonumber
\end{eqnarray}
Where we have introduced the $U(1)_Y$, $SU(2)_L$, and $SU(3)_C$ couplings $g_1$, $g_2$, and $g_3$ respectively, $n_g$ is the number of active generations at the relevant energy scale, the operators corresponding to the wilson coefficients $c_{W}$, $c_{HB}$ and $c_{HW}$ are given by,
\begin{eqnarray}
Q_W&=&\epsilon^{ijk}W^{i,\nu}_\mu W_\nu^{j,\rho}W_\rho^{k,\mu} \, ,\nonumber\\
Q_{HB}&=&(H^\dagger H)B_{\mu\nu}B^{\mu\nu} \, ,\\
Q_{HW}&=&(H^\dagger H)W^i_{\mu\nu}W^{i,\mu\nu} \, ,\nonumber
\end{eqnarray}
and  ``$\cdots$'' represents other operators not generated at tree-level in our EFTs. The final line of Eq.~\ref{eq:anomdimmatrix} is included to indicate that $c_{HWB}$ is not generated at 1-loop by operator mixing and therefore must be generated at two- or higher loop order.
 However, the $T$-parameter is generated at tree-level by the triplet models, and one-loop by any theory which induces $c_{H\Box}$ (namely all but the 2HDM).
In the quartet models, since the only dimension-six operator is the $H^6$ operator, there is no contribution to $S$ and $T$ from the $H^6$ operator. However,  
the $T$-parameter can be generated at tree-level by dimension-eight operators.

Including both the one-loop and running effects we have for the $S$ and $T$ parameters (see e.g. \cite{Alam:1997nk} and \cite{Corbett:2012ja}):
\begin{eqnarray}
\alpha\Delta S&=&2s_{\theta_W}c_{\theta_{W}}v^2c_{HWB}-\frac{1}{6}\frac{e^2}{16\pi^2}\left[4v^2c_{H\Box}\log\left(\frac{M^2}{m_H^2}\right)+\cdots\right],\\
\alpha\Delta T&=&-\frac{1}{2}v^2c_{HD}+\frac{3}{4c^2}\frac{e^2}{16\pi^2}\left[2v^2c_{H\Box}\log\left(\frac{M^2}{m_H^2}\right)+\cdots\right].
\end{eqnarray}
Again we have used $\cdots$ to represent operators generated at higher loop order in our theories.
For the quadruplet models, the dimension-eight operators generate the following $T$-parameter
\bea
	\alpha\Delta T& {\simeq }& -\frac{v^4}{4} c_{T8},
\eea
where we have defined the Wilson coefficient $c_{T8}$ to be the coefficient of the $T$-parameter operator at dimension-eight. This coefficient $c_{T8}$ is then given by,
\begin{equation}
c_{T8}=\frac{2|\lambda_{H3\Phi}|^2}{M^4}\ \ \ \ \&\ \ \ \ c_{T8}=\frac{6|\lambda_{H3\Phi}|^2}{M^4}\, ,
\end{equation}
 for the $Y=Y_H$ and $Y=3Y_H$ quadruplet models respectively.  Note that the coefficients $c_{T8}$ depend on the same quadruplet parameters as the operator $H^6$, and therefore the Wilson coefficient of $H^6$ is also strongly constrained through this correlation.

From GFitter~\cite{Baak:2012kk} we have the central values of the $S$ and $T$ parameters  with correlation matrix $\rho$ as follows,
\bea
	\left(\begin{array}{c}
	S \\
	T 
	\end{array}\right) 
	= \left(\begin{array}{c}
	0.06 \pm  0.09 \\
	0.10   \pm  0.07  
	\end{array}\right), 
 	\quad
 	\rho = \left(\begin{array}{ccc}
 		1.00 & 0.91  \\
 		0.91 & 1.00  
 		\end{array}\right).
\eea
When considering all of the operators discussed above one may perform a sophisticated fit to the EWPD of the many operator coefficients (see e.g. \cite{Falkowski:2014tna}), however for our study we need only consider $c_{H\Box}$ and $c_{HD}$ as discussed above.  Therefore performing a simplified chi-square fit relevant to our EFTs, we obtain constraints on the Wilson coefficients $(c_{HD}, c_{H\Box})$:
\bea
	\left(\begin{array}{c}
	v^2 c_{HD} \\
	v^2  c_{H\Box}  
	\end{array}\right) 
	= \left(\begin{array}{c}
	-0.003654 \pm  0.002677 \\
	8.935   \pm  9.086
	\end{array}\right), 
 	\quad
 	\rho = \left(\begin{array}{cc}
		 1.00  & -0.97 \\
		 -0.97  & 1.00 
 		\end{array}\right).
\eea
We note that  $c_{HD}$ is tightly constrained while $c_{H\Box} $ is not as its contribution to $S$ and $T$ is generated at one-loop.

\subsection{Higgs Diphoton Rate}

In Section II, only the leading tree-level effective operators are written when integrating out the heavy scalars. 
The leading effective operators which contribute to the Higgs diphoton signature are not 
included in our framework as they originate from the one-loop contributions. However because of the precision of the $H\to\gamma\gamma$ measurements we will include them in this section. Note that after integrating out the heavy scalars at one loop one may expect contributions to the $H\to\gamma\gamma$ coupling from the following gauge-invariant dimension-six operators,
\begin{equation}
\mathcal{L}_{H\gamma\gamma}=  c_{HB}(H^\dagger H)B^{\mu\nu}B_{\mu\nu}+  c_{HW}(H^\dagger H)W^{i,\mu\nu}W_{\mu\nu}^i+  c_{HWB}(H^\dagger \tau^i H)W^i_{\mu\nu}B^{\mu\nu}
\end{equation}
However, since we are only interested in the diphoton rate, and not in corrections to the $h\to ZZ$ and $h\to WW$ rates we may simplify the calculation of the Wilson coefficients by only considering one effective operator in the broken phase:
\begin{equation}
\mathcal{L}_{H\gamma\gamma}\to \frac{\alpha }{4\pi } c_{\gamma\gamma}\frac{h}{2v}F_{\mu\nu}F^{\mu\nu}
\end{equation}

The general Higgs diphoton Wilson coefficient $c_{\gamma\gamma}$ for new scalars and fermions at one loop may be found in, e.g. \cite{Djouadi:2005gj}. For the UV complete models considered in Section~\ref{sec:model} we find the wilson coefficients in Table~\ref{tab:cgaga}. As mentioned in the previous section the Wilson coefficients of the Quadruplet model are all proportional to the parameters contributing to the $T$-parameter. As such we will not consider the Quadruplet models for the rest of this section.

\begin{table}
\bgroup
\def\arraystretch{1.8}
\begin{tabular}{|r|c|}
\hline
Model&$c_{\gamma\gamma}$\\
\hline
\hline
($\mathbb{R}$ \& $\mathbb{C}$) Singlet & 0 \\
\hline
2HDM&$ \frac{v^2}{2M^2} Z_3\left(\frac{1}{3}+\frac{2 m_H^2}{45 M^2}\right)$\\
\hline
$\mathbb{R}$ Triplet (Y=0)&$ \frac{v^2}{4M^2}\lambda_{H\Phi}\left(\frac{1}{3}+\frac{2 m_H^2}{45 M^2}\right)$\\
\hline
$\mathbb{C}$ Triplet (Y=-1)&$ \frac{v^2}{4M^2}\left(5\lambda_{H\Phi}+\frac{\lambda'}{2}\right)\left(\frac{1}{3}+\frac{2 m_H^2}{45 M^2}\right)$\\
\hline
\end{tabular}
\egroup
\caption{Wilson coefficient $c_{\gamma\gamma}$ for each UV Complete model in Section~\ref{sec:model}.}\label{tab:cgaga}
\end{table}

Finally the diphoton rate relevant to our models is,
\bea
\label{eq:s-dec1}
	\Gamma(h \to \gamma \gamma) &=&  \frac{\alpha^2 G_F m_h^3}{128\sqrt2 \pi^3} |c_{\gamma\gamma}^{\delta {\rm SM}}+c_{\gamma\gamma}|^2\, ,
\eea
Where we have defined 
\bea
c_{\gamma\gamma}^{\delta {\rm SM}} = \sum_{f=t,b,\tau}N_{c,f} Q_f^2 A_{1/2}(\tau_f) + A_{1}(\tau_W) ,
\eea
as the SM part of the $h\to\gamma\gamma$ width taking into account shifts in the couplings of the Higgs to the $t$-quark and $W$-bosons due to the effective Lagrangian of Eq.~\ref{eq:lorentzlagrangian}.
Here the loop functions $A_{1/2}(\tau)$ and $A_{1}(\tau) $ are defined in Ref.~\cite{Djouadi:2005gj}.

\subsection{Higgs Global Fits}

The Run-I Higgs measurements~\cite{Aad:2012tfa, Chatrchyan:2012xdj, ATLAS:2014yka, Khachatryan:2014jba} provide constraints on some Wilson coefficients in the effective Lagrangian. 
For convenience we reproduce our effective Lagrangian below EWSB here:
\begin{eqnarray}
\mathcal{L}&=&g_{HZZ}^{(3)}hZ_\mu Z^\mu
+g_{HWW}h W_\mu^+W^{-\mu}
+g_{HHH}^{(1)}h^3+g_{HHH}^{(2)}h(\partial_\mu h)(\partial^\mu h)\nonumber\\
&&+\left(g_{He}h\bar e_Le_R+g_{Hu}h \bar u_Lu_R+g_{Hd}h \bar d_L d_R+h.c.\right)+c_{\gamma\gamma}\frac{h}{2v}F_{\mu\nu}F^{\mu\nu}\, .\label{eq:lorentzlagrangian2}
\end{eqnarray}
The corresponding Wilson coefficient dependence can be found in Eq.~\ref{eq:lorentzcoeff} while the Wilson coefficients for each model can be found in Tables~\ref{tab:h4renorm},~\ref{tab:EFTtable},~and~\ref{tab:cgaga}.
We note that the modified Yukawa coupling of the top-quark also causes a shift the Higgs-digluon effective coupling which we have taken into account in our analyses.

\begin{figure}[h]
 \includegraphics[width=0.5\textwidth]{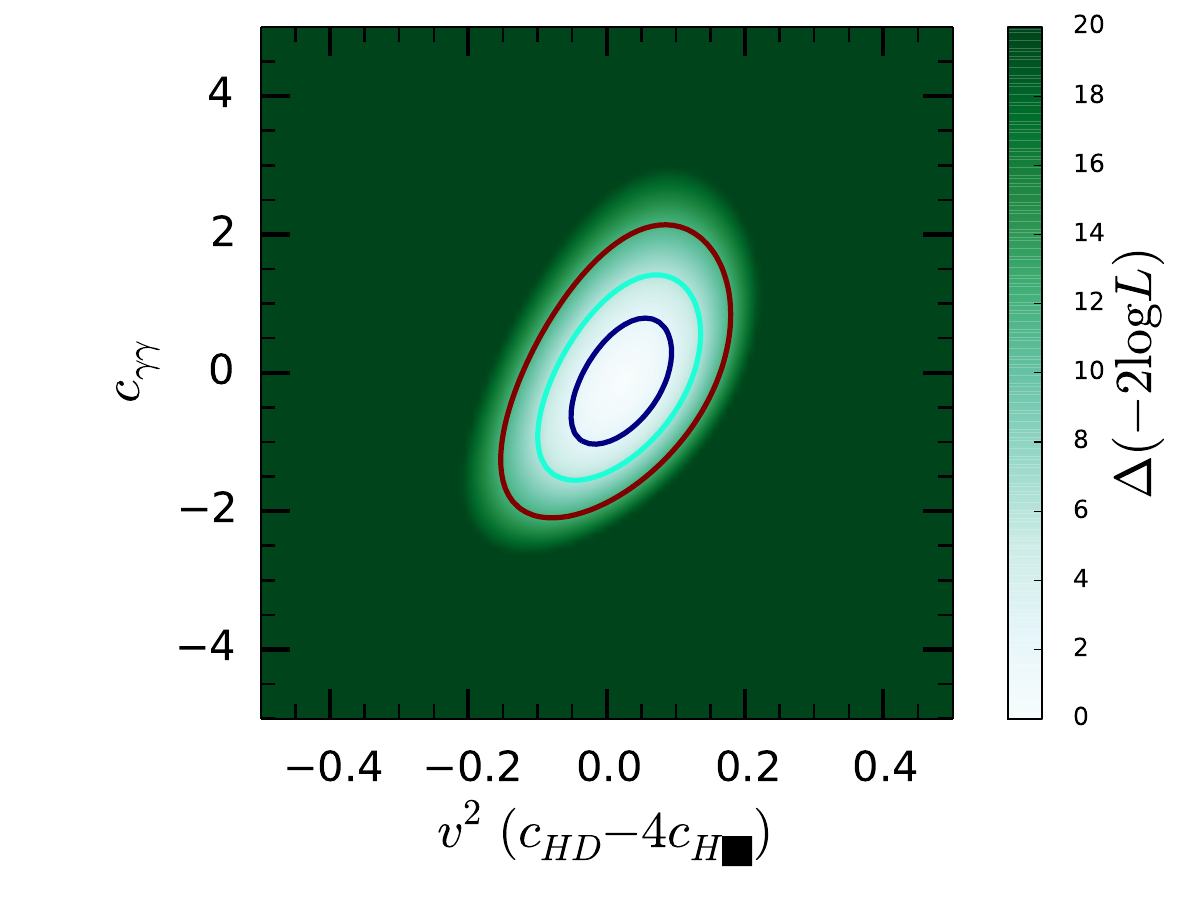} 
 \caption{The $1$, $2$, and $3\ \sigma$ level profiled contours between $v^2(c_{HD} - 4 c_{H\Box})$ and $c_{\gamma \gamma}$, given that other operators are fixed to be the local best values.}\label{fig:corr,hboxgaga}
\end{figure}

These Wilson coefficients contribute to the Higgs signal strengths $\mu = \frac{\sigma \times A \times \epsilon}{[\sigma \times A \times \epsilon]^{\rm SM}}$ extracted from the Higgs coupling data, where $A \times \epsilon$ is the product of the acceptance and the efficiency. 
Since the Higgs discovery global fits to the effective operators relevant to Higgs physics have become an important area of research \cite{Corbett:2012ja,Corbett:2012dm,Corbett:2013pja} and recently they have gone beyond simple inclusion of signal strengths to inclusion of kinematic variables and off-shell measurements \cite{Corbett:2015ksa,Butter:2016cvz}. They have also been considered in scenarios where EWSB is not linearly realized \cite{Brivio:2013pma,Corbett:2015mqf,Brivio:2016fzo}. However for the sake of our analyses we require a much smaller set of effective operators, therefore we perform a simplified global fit to the Higgs signal strengths $\mu_i$ using the program Lilith~\cite{Bernon:2015hsa}. 

In Lilith, all the Run I LHC Higgs measurements~\cite{Aad:2012tfa, Chatrchyan:2012xdj, ATLAS:2014yka, Khachatryan:2014jba} are taken into account, and a likelihood statistical procedure is performed to obtain the 
constraints on the signal strengths. 
It is based on the assumption that the Higgs measurements are approximately Guassian and thus the likelihood function $L(\mu)$ could be simply reconstructed. Under this assumption adapted by Lilith, the
$-2 \log L(\mu)$ follows a $\chi^2$ law for each observable,
\bea
	- 2 \log L(\mu_i) = \left( \frac{\mu_i - \hat{\mu}_i } {\Delta \mu_i}\right)^2,
\eea
where $\hat{\mu}_i$ is the theoretical prediction of the measured Higgs signal strengths $\mu_i$ with Gaussian uncertainty $\Delta \mu_i$. 
The full likelihood $L(\mu) = \prod_{i} L(\mu_i)$ is defined as
\bea
	- 2 \log L({\bf{\mu}}) = \chi^2({\bf{\mu}}) = ({\bf{\mu}} - \hat{\bf{\mu}})^T C^{-1} ({\bf{\mu}} - \hat{\bf{\mu}}),
\eea
where $C^{-1}$ is the inverse of the $n \times n$ covariance matrix, with $C_{ij} = \textrm{cov}[\hat{\mu}_i, \hat{\mu}_j]$.

Then the constraints on the signal strengths are recast as bounds on the Wilson coefficients. 
We perform a global fit on these Wilson coefficients $(c_{HD}, c_{H\Box}, c_{\gamma\gamma}, c_{iH})$ with $i = t,b,\tau$,
and then project our results into the sub-space in each scalar model.
First we perform the six-parameter fit, and obtain
\bea
	\left(\begin{array}{c}
	v^2 * c_{tH} \\
	v^2 * c_{bH} \\
	v^2 * c_{\tau H} \\
	v^2 * c_{HD} \\
	v^2 * c_{H\Box} \\
	c_{\gamma\gamma}  
	\end{array}\right) 
	= \left(\begin{array}{c} 
	-0.02224 \pm  0.4609   \\
	-0.111   \pm  0.5933   \\
	 0.02993 \pm  0.4859   \\
	 0.1399  \pm  0.6514   \\
	 0.02283 \pm  0.2255   \\
	-0.3373  \pm  2.028    \\
	\end{array}\right), 
 	\quad
 	\rho = \left(\begin{array}{cccccc} 
 1.00  & 0.60  & 0.40  & 0.21  & -0.26 & -0.48 \\
 0.60  & 1.00  & 0.38  & 0.19  & 0.43  & -0.47 \\
 0.40  & 0.38  & 1.00  & 0.29  & -0.11 & -0.46 \\
 0.21  & 0.19  & 0.29  & 1.00  & 0.19  & -0.39 \\
 -0.26 & 0.43  & -0.11 & 0.19  & 1.00  & 0.16  \\
 -0.48 & -0.47 & -0.46 & -0.39 & 0.16  & 1.00 
 		\end{array}\right),
\eea
where $\rho$ is the correlation matrix for this global fit. 
These Wilson coefficients are typically small due to suppression by $\frac{v^2}{M^2}$. 
However from Subsection~\ref{subsec:ewpmeasurements} we know we must also consider the EWSB constraints. Assuming equal weight and combining with the constraints coming from the $S$ and $T$ parameters, 
we find that $C_{HD}$ is very tightly constrained:
\bea\label{eq:Higgs_EW_constraints}
	\left(\begin{array}{c}
	v^2 * c_{tH} \\
	v^2 * c_{bH} \\
	v^2 * c_{\tau H} \\
	v^2 * c_{HD} \\
	v^2 * c_{H\Box} \\
	c_{\gamma\gamma} 
	\end{array}\right) 
	= \left(\begin{array}{c}
  -0.04967   \pm  0.4551     \\
  -0.121     \pm  0.5917     \\
  -0.003816  \pm  0.4722     \\
  -0.0004666 \pm  0.0003861  \\
   0.02302   \pm  0.2184     \\
  -0.1513    \pm  1.891      \\
	\end{array}\right), 
 	\quad
 	\rho = \left(\begin{array}{cccccc}
1.00  & 0.58  & 0.35  & 0.07  & -0.32 & -0.43 \\
0.58  & 1.00  & 0.35  & -0.08 &  0.39 & -0.44 \\
0.35  & 0.35  & 1.00  & 0.04  & -0.18 & -0.40 \\
0.07  & -0.08 & 0.04  & 1.00  & -0.20 & -0.05 \\
-0.32 & 0.39  & -0.18 & -0.20 & 1.00  & 0.27  \\
-0.43 & -0.44 & -0.40 & -0.05 & 0.27  & 1.00 
 		\end{array}\right).
\eea
We also obtain that $v^2 * (c_{HD} - 4c_{H\Box}) =  -0.09256 \pm 0.8731$, which by Eq.~\ref{eq:lorentzcoeff} we see is a very important constraint on both the momentum dependent and momentum independent tri-higgs couplings. 
In Figure~\ref{fig:corr,hboxgaga} we show the $v^2(c_{HD} - 4 c_{H\Box}) \times c_{\gamma \gamma}^{\rm scalar}$ plane where we have marginalized over the parameters not shown.

We see from Figure~\ref{fig:corr,hboxgaga} that the independent constraint on $c_{\gamma\gamma}$ provides an important constraint in the space of Wilson coefficients which will translate to a constraint on the various four scalar couplings of the UV models and therefore through their correlation with the Wilson coefficient $c_{H}$ on the affects of the $Q_H$ operator. We project these constraints in the EFT framework onto the UV complete model parameters in the next subsection.

\subsection{Implications for the UV Physics}
\begin{figure}[h]
\begin{tabular}{cc}
 \includegraphics[width=0.32\textwidth]{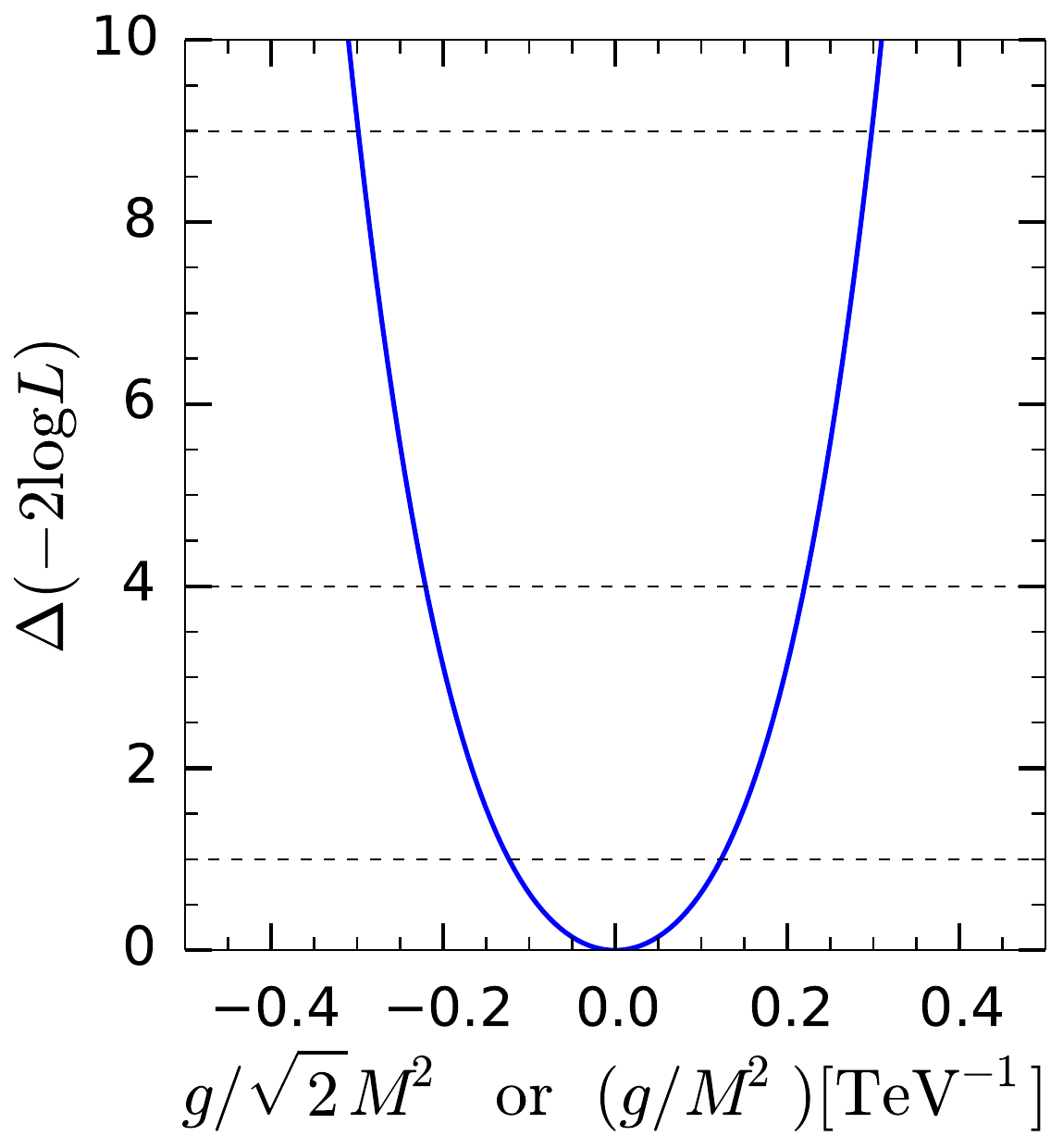} &
 \includegraphics[width=0.45\textwidth]{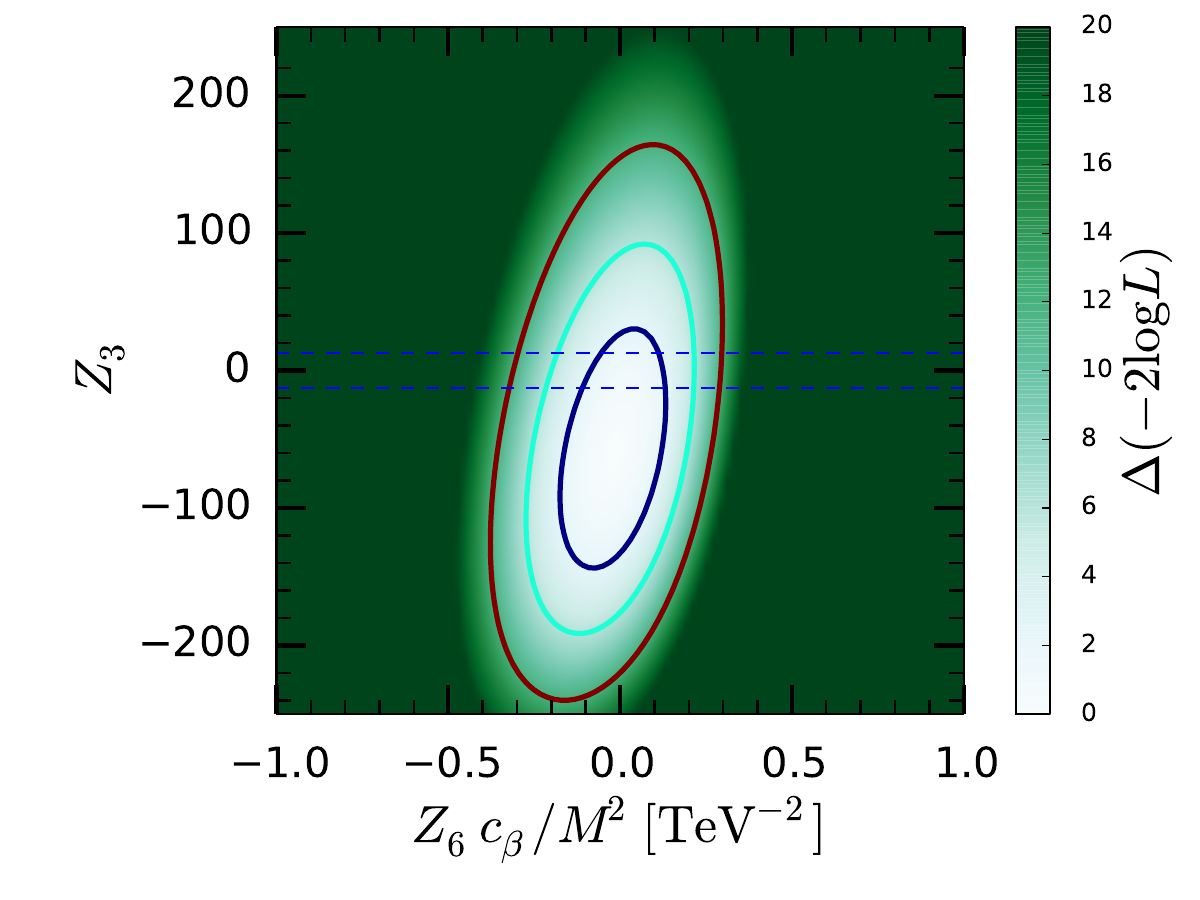} \\
 \includegraphics[width=0.45\textwidth]{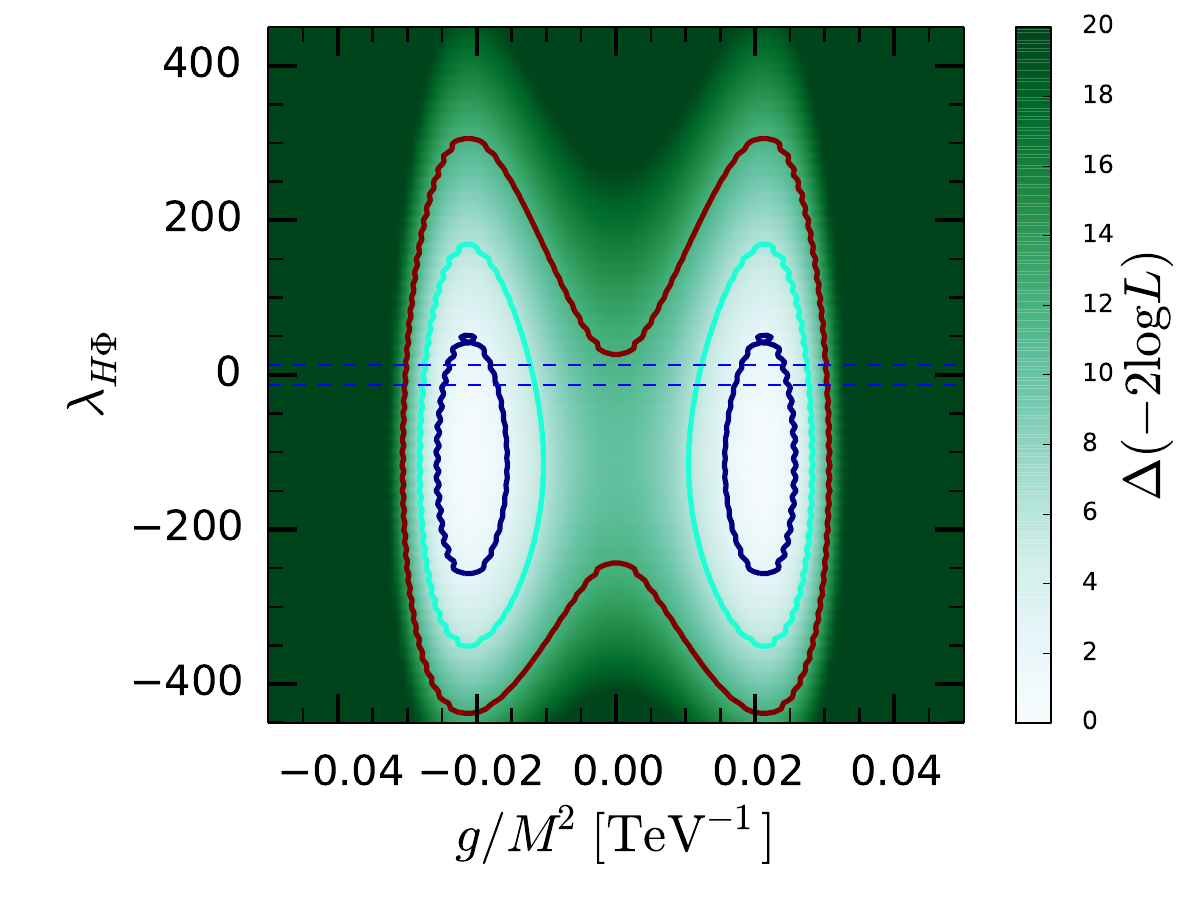} &
 \includegraphics[width=0.45\textwidth]{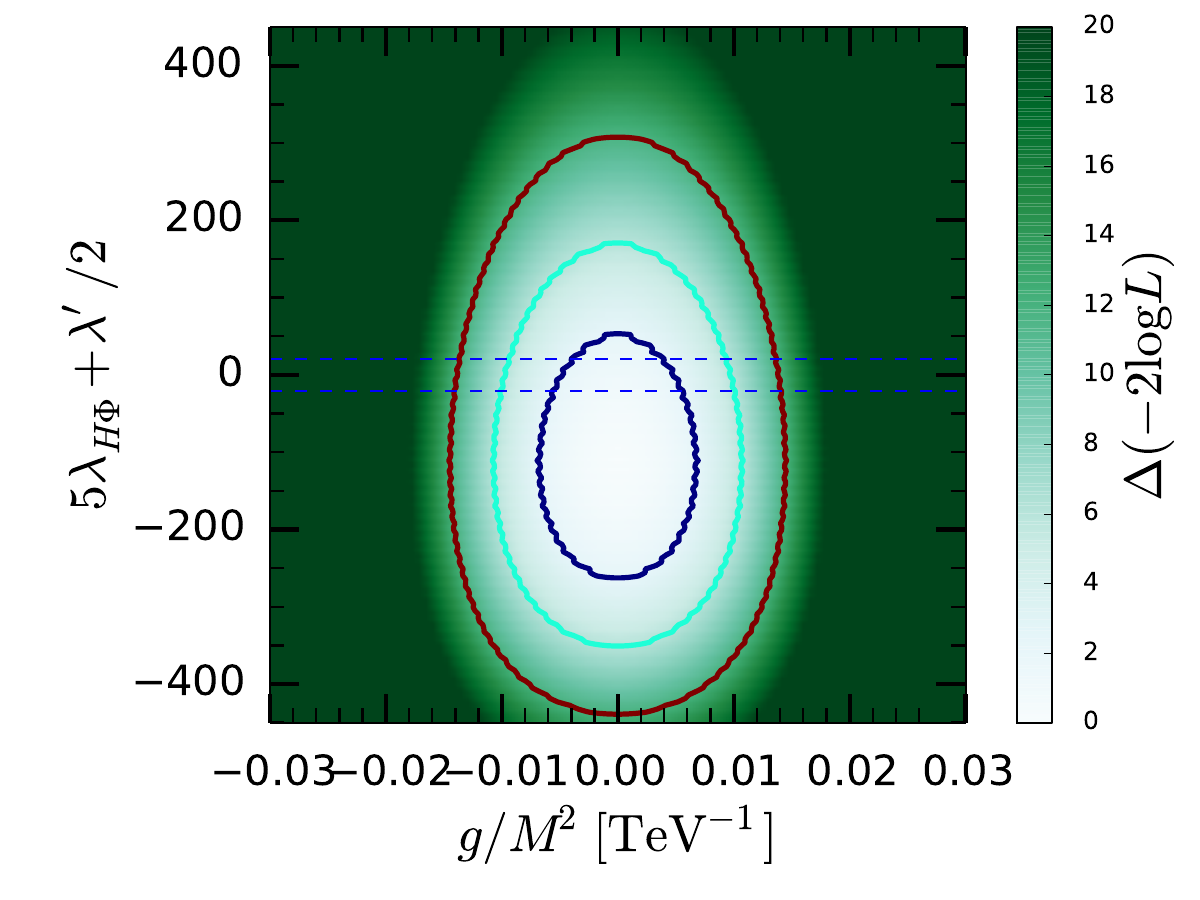} 
 \end{tabular}
 \caption{In the upper left panel, the log likelihood vs the coupling $g/\sqrt2 M^2$ ($g/M^2$) in the real (complex) singlet model. In the others, we show the $1$, $2$, and $3\ \sigma$ contours on the model parameters in the Type-I 2HDM (top right), the real triplet (bottom left) and complex triplet model (bottom right). The colored contours show the log likelihood values in the global fit. The blue dashed lines denotes the perturbativity bounds of the dimensionless scalar couplings: $\pm 4\pi$. 
 %In all models, we  use a benchmark value  $M = 1 $ TeV. 
 }
 \label{fig:modelfitss}
\end{figure}

\begin{table}
\bgroup
\def\arraystretch{1.8}
\begin{tabular}{|r|c|c|}
\hline
\hline
Model& $Z_6/M^2$ or $g/M^2$ or $\lambda_{H3\Phi}/M^2$ &  $Z_3$ or $\lambda_{H\Phi}$ \\
\hline
($\mathbb{R}$ \& $\mathbb{C}$) Singlet & $g/\sqrt2 M^2$  or $g/M^2$= $ 0.00   \pm  0.131 $ TeV$^{-1}$ & N/A\\
\hline
2HDM&$ 0.016  \pm  0.144 $ TeV$^{-2}$ & $ -4\pi   \pm  25.29 $  \\
\hline
$\mathbb{R}$ Triplet (Y=0)&$ -0.03 \pm  0.007 $ TeV$^{-1}$ & $ -4\pi  \pm  19.97$\\
\hline
$\mathbb{C}$ Triplet (Y=-1)&$0    \pm  0.0071$ TeV$^{-1}$ & $5\lambda_{H\Phi}+ \lambda'/2$  = $-22\pi    \pm  141.3$\\
\hline
$\mathbb{C}$ Quadruplet ($Y=1/2$ \& $Y=3/2$)&  $\lambda_{H3\Phi}/M^2$  or $3\lambda_{H3\Phi}/M^2$= $ 0.00   \pm  0.053 $ TeV$^{-2}$ & N/A \\
\hline

%\hline
%Model& $Z_6$ or $g/v$ or $\lambda_{H3\Phi}$ &  $Z_3$ or $\lambda_{H\Phi}$ \\
%\hline
%($\mathbb{R}$ \& $\mathbb{C}$) Singlet & $g/\sqrt2 v$  or $g/v$= $ 0.00   \pm  2.344 $ & N/A\\
%\hline
%2HDM&$ 0.2659  \pm  2.365 $ & $ -4\pi   \pm  25.29 $\\
%\hline
%$\mathbb{R}$ Triplet (Y=0)&$ -0.5005  \pm  0.1179 $ & $ -4\pi  \pm  19.97$\\
%\hline
%$\mathbb{C}$ Triplet (Y=-1)&$0    \pm  0.118$ & $5\lambda_{H\Phi}+ \lambda'/2$  = $-22\pi    \pm  141.3$\\
%\hline
%$\mathbb{C}$ Quadruplet ($Y=1/2$ \& $Y=3/2$)&  $\lambda_{H3\Phi}$  or $3\lambda_{H3\Phi}$= $ 0.00   \pm  0.86 $ & N/A \\
%\hline
\end{tabular}
\egroup
\caption{The central values and $1\sigma$ errors of the model parameters for each UV complete model. 
%In all models, we  use a benchmark value  $M = 1 $ TeV. 
We also limit the range of the dimensionless Higgs couplings to be less than $\pm 4\pi$.}\label{tab:modeluv}
\end{table}

In the global fitting procedure, all the Wilson coefficients are assumed to be independent. 
We know from Section~\ref{sec:model} that in the specific scalar extended models some Wilson coefficients are correlated and some Wilson coefficients may be absent altogether. 
These correlations and absences may be seen in Table~\ref{tab:EFTtable}. 
Therefore, it proves useful to recast the global fit results to obtain constraints on the UV model parameters in each model.

We perform the global fit using the Lilith program in each scalar extended model. 
In Fig.~\ref{fig:modelfitss}, we show the $1$, $2$, and $3\sigma$ contours on the model parameters in the real and complex singlet, Type-I doublet, and complex/real triplet models.
%, with the benchmark value $M = 1 $ TeV assumed. 
At the same time, we also show the central values and errors for the model parameters in Tab.~\ref{tab:modeluv}.
These plots exhibit similar features. 
First, the Higgs-Higgs-scalar coupling $g/M^2$ or $Z_6/M^2$ is constrained to be $\mathcal{O}(0.1 - 1)$ by 
the Higgs gauge boson couplings in the singlet and doublet models, while in the triplet models   the $T$-parameter puts tighter constraints on the parameter $g/M^2$. 
Secondly, for the doublet and triplets,  the Higgs to diphoton rate puts additional constraints on the couplings which 
contribute to the $c_{\gamma\gamma}$. Converting 
to the couplings in the UV model, we are not further able to constrain the Higgs-Higgs-scalar-scalar couplings of the triplet models $\lambda_{H\Phi}$ and $\lambda'$, because the constraints shown in Fig.~\ref{fig:modelfitss} and Tab.~\ref{tab:modeluv} are very loose. Even the perturbativity constraint, shown as the blue dashed lines in Fig.~\ref{fig:modelfitss},  is tighter than the constraint from the global fit. So to place constraints on the Wilson coefficients of $Q_H$ for the 2HDM and triplet models, we have to rely on di-Higgs collider constraints. 
Finally, we note that although the global fit cannot constrain the renormalizable Higgs self coupling $\lambda$, 
it is able to constrain the dependence of the $h(\partial h)^2$ effective coupling indirectly. We have neglected to project our global fit into the parameter space of the quadruplet as it is so strongly constrained by the $T$-parameter and the triplet serves as an example of the affects.

While these indirect constraints on the UV models from the global fit are interesting and useful for our di-Higgs analysis in the following section, stronger constraints may of course be found in UV complete considerations of these models. The ability to loosely constrain numerous models at once from simple Higgs global fits is nonetheless intriguing and (especially in the advent of a significant deviation from the SM expectation) a useful way to direct UV complete searches of greater depth in the future.
%

%%%%%%%%%%%%%%%%%%%%%%%%%%%%%%%%%%%%%%%%%%%%%%%%%%%%%%%%%%%%%%
\section{Di-Higgs Production at the 100 TeV Collider}\label{sec:dihiggs}

The measurement of the triple Higgs coupling using non-resonant di-Higgs production at both the LHC and future 100 TeV collider has been studied in great detail in the literature which was recently reviewed in~\cite{Contino:2016spe}. 
Among all the channels for the Higgs decay final state, the $bb\gamma\gamma$ channel~\cite{Barger:2013jfa,Dolan:2012rv,He:2015spf,Huang:2016cjm,Barr:2014sga,Kling:2016lay,Huang:2015tdv,Cao:2016zob} is the most promising due to the combination of large $h\to bb$ branching ratio and more accurate reconstruction of photon momentum compared with other channels which helps reduce the backgrounds. 

Three different topologies of Feynman diagrams of the $pp \to hh $ process via the gluon fusion production are shown in Fig.~\ref{fig:ggtohh}. Due to the destructive interference between the triangle and box diagram for the di-Higgs production in the gluon fusion channel, it is believed that at 14 TeV LHC with 3 ${\rm ab}^{-1}$ luminosity, the triple Higgs coupling $-g^{(1)}_{HHH}/\lambda_{SM}v$ would be constrained to only $[-0.8,7.7]$ at 95\% CL~\cite{ATL-PHYS-PUB-2017-001}.
In all models considered in this article, the Wilson coefficients of the $|H|^6$ operator cannot be chosen arbitrarily large. Based on the considerations of the validity of EFT and perturbative constraints, we estimate the value of the modified trilinear Higgs coupling to be within the range $(-0.1 \lambda_{\rm SM}, 2\lambda_{\rm SM})$,
and take the cutoff scale to be $2$ TeV. The higher  the cutoff scale, we expect the narrower range of the trilinear Higgs coupling.
%~\footnote{} 
On the other hand, at a 100 TeV collider with 30~${\rm ab}^{-1}$ luminosity, the SM value of the triple Higgs coupling can be measured with around 10\% uncertainty~\cite{Contino:2016spe}, and even around 4\% based on the latest study~\cite{slides}. Therefore, we expect that 100 TeV collider provides a good opportunity to explore the Wilson coefficients $c_H$ in various models we have considered\footnote{Though in 2HDM the modification of the Wilson coefficient $c_{tH}$ (not yet constrained tightly) can be large enough to modify the di-Higgs production cross section to give some evidence in 14 TeV LHC, yet other models we considered definitely need the help of 100 TeV collider to probe, due to small
or zero $c_{tH}$. 
%In this work we focus on modification of the Wilson coefficient $c_H$ instead of $c_{tH}$, thus we only focus on 100 TeV collider. 
}.

%{\color{green} In all the models considered in this article, the Wilson coefficients of the $h^6$ operator cannot be chosen arbitrarily large. Based on the perturbative constraints, in all considered models the Higgs self coupling is typically limited to be around $(-0.1 \lambda_{\rm SM}, 2\lambda_{\rm SM})$, which is really hard to be probed at the high luminosity LHC runs. On the other hand, at a 100 TeV collider with 30 ~${\rm ab}^{-1}$ luminosity, the SM value of the triple Higgs coupling can be measured with around 10\% uncertainty~\cite{Contino:2016spe}, and even around 4\% based on the latest study~\cite{slides}.
%be discovered with a significance of $S/\sqrt{B} =8$ ~\cite{Barr:2014sga} with a very well constructed di-photon invariant mass. 
%Therefore, we expect that 100 TeV collider provides a good chance to explore the Wilson coefficients $c_H$ in various models we have considered. }

\subsection{General Formalism on Di--Higgs Production}

In our EFT framework, the effective Lagrangian relevant to the di-Higgs production is 
\begin{eqnarray}
\mathcal{L}&=&
 g_{HHH}^{(1)}h^3+g_{HHH}^{(2)}h(\partial_\mu h)(\partial^\mu h) +\left( g_{tH}h \bar t_Lt_R+g_{bH}h \bar b_L b_R+g_{HHt}hh\bar t_L t_R + g_{HHb}hh\bar b_L b_R +h.c.\right) \, ,\label{eq:higgsselflagrangian}
\end{eqnarray}
where 
\begin{eqnarray}
g_{HHH}^{(1)}&=& -  {\lambda_{SM} v}\left[1-\frac{v^2}{4}(c_{HD}-4c_{H\Box}+{\frac{4}{\lambda_{RF}}c_H})\right],		\label{eq:higgsselfcoeff}\\
g_{HHH}^{(2)}&=& v(c_{HD}-4c_{H\Box}) ,	\\
	g_{\psi H} &=& -\frac{m_\psi}{v}\left[1-\frac{v^2}{4}(c_{HD}-4c_{H\square})\right]+\frac{c_{\psi H}v^2}{\sqrt{2}},\label{eq:gpsiH} \\
g_{HH\psi} &=& \frac{3c_{\psi}v}{2\sqrt{2}} \label{eq:gpsiHH}
\end{eqnarray}
with the SM vacuum expectation value $v \equiv \frac{1}{2(\sqrt{2}G_F)^{1/2}}$ and the SM dimensionless coupling $\lambda_{SM}  \equiv \sqrt{2}G_F {m_H^2} $. 
From the above Lagrangian, we note that in the Warsaw basis, in addition to the SM trihiggs couplings, we also have derivative triple-Higgs couplings, which may contribute differently to the distribution compared with solely non-derivative couplings.

\begin{figure}
\centering     
\subfigure[ \tiny{gluon fusion box}]{\includegraphics[width=60mm]{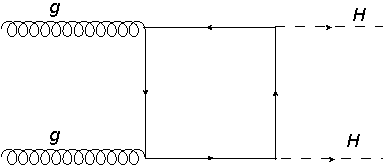}}
\subfigure[\tiny{gluon fusion triangle}]{\includegraphics[width=60mm]{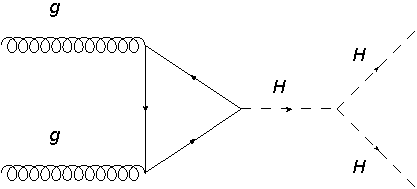}}
\subfigure[\tiny{gluon fusion tthh}]{\includegraphics[width=60mm]{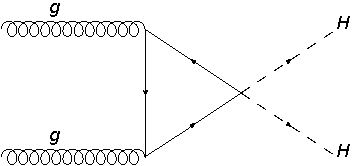}}
\caption{Different topologies of the $gg \to hh $ process via the gluon fusion production.}
\label{fig:ggtohh}
\end{figure}

According to  Fig.~\ref{fig:ggtohh}, the parton amplitude of the di-Higgs production $g(p_1)g(p_2)\to h(p_3)h(p_4)$ via the gluon fusion process is
\begin{eqnarray}
{\cal M}_{hh}&=&-\frac{\alpha_s\hat{s}\delta^{ab}}{4\pi v^2}\epsilon^a_{\mu}(p_1)\epsilon^b_{\mu}(p_2)\left\lbrace  \left[\left(\frac{g_{Ht}v}{m_t}\frac{g^{(1)}_{HHH}}{v}\frac{3m_H^2}{\hat{s}-m_H^2}-g^{(2)}_{HHH}v\frac{\hat{s}+2m_H^2}{\hat{s}-m_H^2} +\frac{2v^2}{m_t}g_{HHt}\right)F_\triangle +\frac{g_{Ht}^2v^2}{m_t^2}F_\Box \right]A^{\mu\nu} \right. \nonumber \\
&&\left. +\frac{g_{Ht}^2v^2}{m_t^2}G_\Box B^{\mu\nu}  \right\rbrace,
\label{eq:hhxseceq}
\end{eqnarray}
where the Lorentz structures are 
\begin{eqnarray}
A^{\mu\nu}&=&g^{\mu\nu}-\frac{p_{{1}}^{\nu}p_{{2}}^{\mu}}
{p_{{1}}\cdot p_{{2}}}, \\
B^{\mu\nu}&=&g^{\mu\nu}+\frac{p_{{3}}^{2}p_{{1}}^{\nu}
p_{{2}}^{\mu}}{p_{T}^2 p_{{1}}\cdot p_{{2}}}-\frac{2p_{{2}}
\cdot p_{{3}}p_{{1}}^{\nu}p_{{3}}^{\mu}}{p_{T}^2 p_{{1}}
\cdot p_{{2}}}-\frac{2p_{{1}}\cdot p_{{3}}p_{{2}}^{\mu}
p_{{3}}^{\nu}}{p_{T}^2 p_{{1}}\cdot p_{{2}}}
+\frac{2p_{{3}}^{\mu}p_{{3}}^{\nu}}{p_{T}^{2}},
\end{eqnarray}
and $F_{\triangle}$, $F_\Box$, and $G_\Box$ are the form factors for triangle and box diagrams   which can be found in  Ref.~\cite{Plehn:1996wb}. 
Correspondingly,  the   differential cross-section   for di-Higgs production  is given by:  
\begin{eqnarray}
\frac{d\sigma(pp\to hh)}{d\hat{s} d\hat{t}} = \frac{1}{S}{\mathcal{L}}_{gg}\left(\frac{\hat{s}}{S},\sqrt{\hat{s}}\right) \frac{{|{\cal M}_{hh}|^2}}{32\pi\hat{s}}, 
\end{eqnarray}
where $S$  is  the  center-of-mass energy squared of the proton-proton system, $\hat{s}=(p_1+p_2)^2$,   $\hat{t}=(p_1-p_3)^2$ and the parton luminosity function is defined as
\bea
{\mathcal{L}_{gg}}(y,\mu_F)=\int ^1_y \frac{dx}{x}f_{g/p}(x,\mu_F)f_{g/p}(\frac{y}{x},\mu_F),
\eea
with $f_{g/p}$  the gluon distribution function, and $\mu_F$  the
factorization scale. 
As we have previously noted, the triangle diagram and box diagram interfere destructively and the smallest cross section is obtained when $g^{(1)}_{HHH}/v\approx -2.5\lambda_{SM}$ assuming no derivative interaction and no corrections to the quark-Higgs couplings. 
Due to this fact, the variation in the gluon fusion to di-Higgs cross section about the SM value of $g^{(1)}_{HHH}=-\lambda_{SM}v$ is not symmetric. When $g^{(1)}_{HHH}$ decreases, the total cross section decreases, till  $g^{(1)}_{HHH}$ reaches $-2.5\lambda_{SM}$. Any further decrease in $g^{(1)}_{HHH}$ results in increasing of the cross section with respect to its minimum value at $g^{(1)}_{HHH}/v\approx -2.5\lambda_{SM}$ eventually surpassing the SM value for $g^{(1)}_{HHH}$ values lower than $-5\lambda_{SM}$. On the other hand as $g^{(1)}_{HHH}$ increases from zero, the total cross section increases.
In our case, the situation is more complicated, we now have both an additional vertex and corrections to the quark Higgs couplings.

\subsection{Di--Higgs Cross Section}

\begin{figure}
	\begin{center}
				\includegraphics[width=0.32\textwidth]{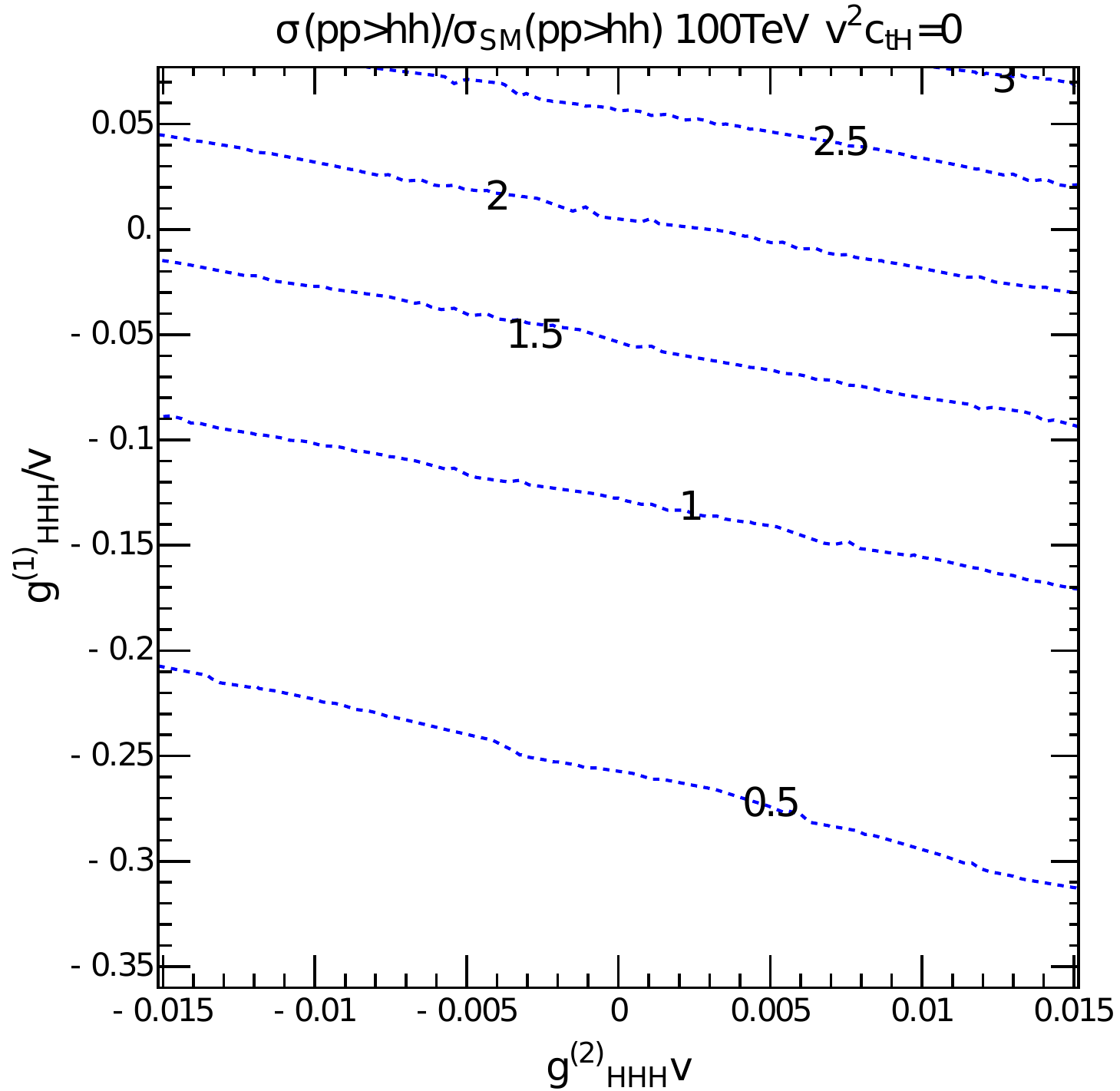}
		\includegraphics[width=0.32\textwidth]{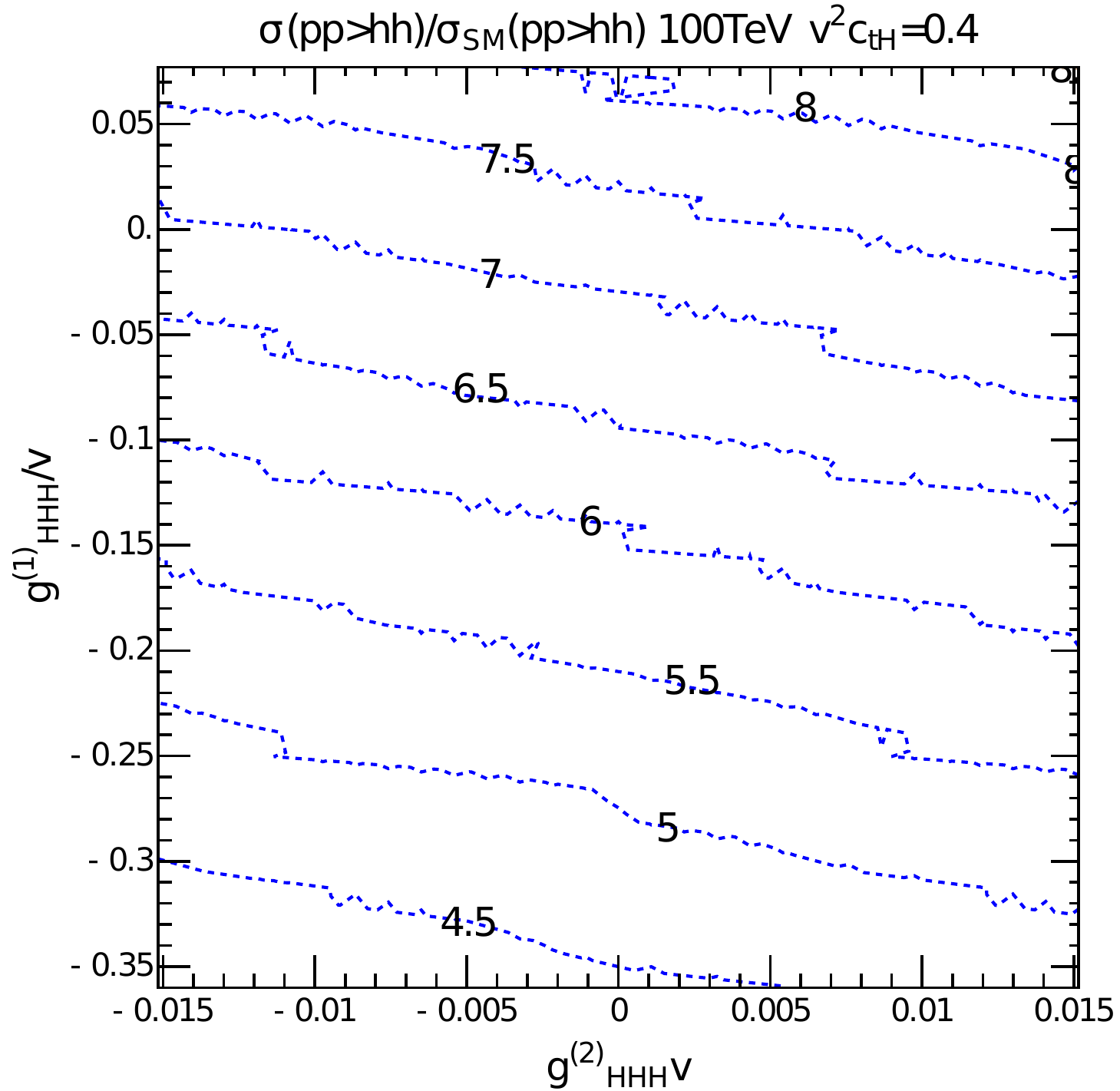}
		\includegraphics[width=0.32\textwidth]{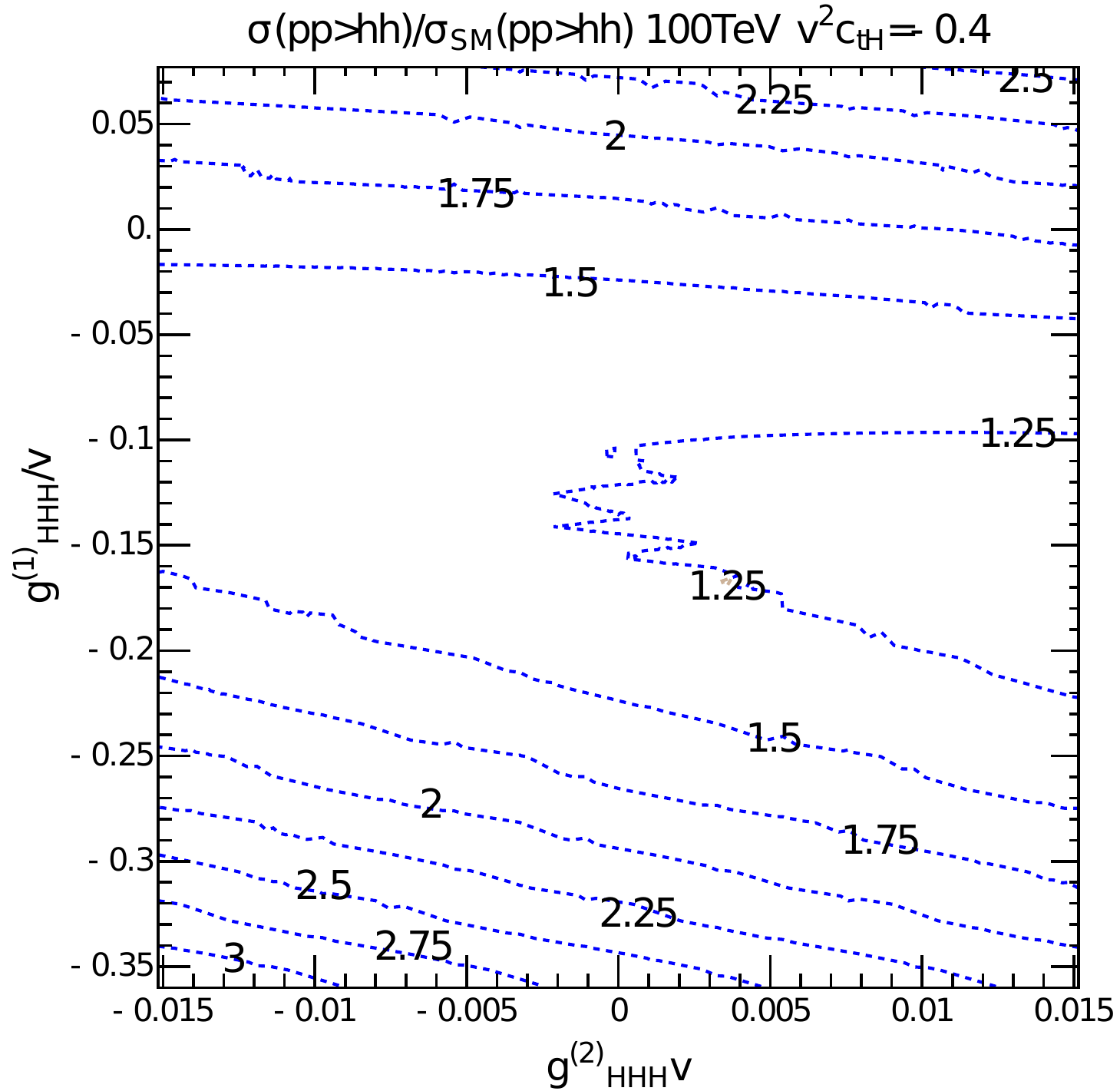}
	\end{center}
	\caption{The ratio of the cross sections of the $pp \rightarrow hh$ process to the SM di-Higgs cross section denoted by the dashed blue contours in the $(g^{(1)}_{HHH}, g^{(2)}_{HHH})$ plane, the plots from left to right correspond to three different value of $c_{tH}={0,\ 0.4,\ -0.4}$. We adopt the NNLL matched NNLO SM di-Higgs cross section: $1.75$ pb~\cite{Contino:2016spe}. }\label{fig:xsect}
\end{figure}

In Figure~\ref{fig:xsect} we show the cross section contours of the $pp \rightarrow hh$ process in the $(g^{(1)}_{HHH}/v, g^{(2)}_{HHH}v)$ plane with three different values of $c_{tH}$. 
To evaluate the range of trihiggs couplings $g^{(1)}_{HHH}/v$ and $g^{(2)}_{HHH}v$, we first use the Eq.~\ref{eq:lorentzcoeff} and Table~\ref{tab:EFTtable} to express the two couplings in terms of the parameters in the UV model, then varies the dimensionless parameters in the UV models within the range $\pm 4 \pi$, couplings with mass dimension to be in the range $\pm$1 TeV, and the cutoff scale are set to be 2 TeV. These values are chosen such that our EFT matching procedure is valid (dimension-eight operators will not be enhanced by the factor $g^2/M^2$) and the contribution of the kinematic region larger than cutoff scale to the total rate is negligible due to the suppression of the parton luminosity. After these consideration, we choose relatively loose ranges for the two couplings: $g^{(1)}_{HHH} \subset (-0.36, 0.07)$ and $g^{(2)}_{HHH} \subset (-0.015, 0.015)$.

%In Figure~\ref{fig:xsect} we show the cross section contours of the $pp \rightarrow hh$ process in the $(g^{(1)}_{HHH}/v, g^{(2)}_{HHH}v)$ plane with three different values of $c_{tH}$. 
%
For $c_{tH}=0$, the anomalous Higgs fermion coupling $g_{HHt}$ in Eq.~\ref{eq:gpsiHH}  vanishes and the corrections to the quark Higgs couplings are proportional to $c_{HD}-4c_{H\square}$. 
In such a case, only the first triangle and box diagrams of Figure~\ref{fig:ggtohh} contribute to the cross section with approximate SM quark Higgs couplings. Hence, one can find that, along the positive vertical direction, given a fixed value of $g^{(2)}_{HHH}$, the cross section increases. Along the $g^{(2)}_{HHH}$ direction, one can find that a positively increasing value of $g^{(2)}_{HHH}$ will lead to an increase in the total cross-section. 
This can be understood from Eq.~\ref{eq:hhxseceq}, where we observe that, with a positive $g^{(2)}_{HHH}$, the second term inside the bracket in front of the $F_\triangle$ which is induced by the derivative interaction will add destructively with the first term which is induced by the ordinary triple Higgs interaction, such that the effect of destructive interference between the box and triangle diagrams is alleviated.

In the case of $c_{tH}=0.4$, the cross section increases significantly when compared with the cross section for $c_{tH}=0$, this can also be understood from Eq.~\ref{eq:hhxseceq} and Eq.~\ref{eq:gpsiH}: The positive $c_{tH}$ will decrease the magnitude of $g_{tH}$ and also gives a new positive term generated by $tthh$ vertex, which will alleviate the destructive interference.
In the case of $c_{tH}= -0.4$, the cross section will reach some minimum value between $g_{HHH}^{(1)}/v = -0.1$ and $-0.15$ due to the destructive interference.
Below the miminum points, for a fixed $g_{HHH}^{(1)}$, increasing $g_{HHH}^{(2)}$ will decrease the cross section, because at this point the amplitude from the triangle diagram becomes dominant, increasing $g_{HHH}^{(2)}$ will decrease the magnitude of the term inside the bracket in front of the $F_\triangle$, thereby decreasing the cross section.

\subsection{Monte Carlo Simulation and Validation}

In order to perform our simulations we begin by using FeynRules~\cite{Alloul:2013bka} to generate an UFO model file adding the effects of the dimension-six operators in Eq.~\ref{eq:higgsselflagrangian}. We then modify the model file to include the full triangle and box form factors as computed in~\cite{Frederix:2014hta}. Then we implement MadGraph 5.2.4.3~\cite{Alwall:2014hca} to generate events. 
We use Pythia 6~\cite{Sjostrand:2006za} for the parton shower and the FCC card in Delphes 3.4~\cite{Selvaggi:2014mya} for simulating the detector. The following analysis is only concerned with statistical uncertainties as the systematical uncertainties are unknown at the moment. When taken into account they will lower the significance levels given in this section.

We refer to the cuts applied while generating the events in MadGraph/Delphes as preselection cuts in the Table~\ref{tab:efficiency}. They are as follows\footnote{For $bb\gamma\gamma$ and $bbj\gamma$ events, we also implement the $50<m_{bb}<250 {\rm GeV}$ and $90<m_{j\gamma,\gamma\gamma}<160 {\rm GeV}$ to increase the efficiency of the sample, and we found that the events outside these cuts contribute negligibly to the final results.}:
\begin{align}\label{eqn:basiccuts}
|\eta_{j,b,\gamma}|<2.5,\ \ \ \ \ \ \Delta R_{jj,j\gamma}>0.4,\, pT_{j,b}>20\, \text{GeV},\ \ \ \ \ \ pT_{\gamma}>10\,\text{GeV}
\end{align}

Important irreducible backgrounds consist of $Z(b\bar{b})h(\gamma\gamma)$, $t\bar{t}h(\gamma\gamma)$, $b\bar{b}h(\gamma\gamma)$, $b\bar{b}\gamma\gamma$ production. Apart from these, there are $bbj\gamma$, $jj\gamma\gamma$, $c\bar{c}\gamma\gamma$ and $bbjj$ channel that can potentially have a contribution to the background. Jet fake rates to photons are taken to be $0.012\%$, while jet and charm mistagging rates to bottom quarks are taken to be $1\%$ and $10\%$ respectively~\cite{deFavereau:2013fsa}. The backgrounds can be greatly reduced by vetoing extra jets, i.e., by demanding exact two $b$-tagged jets in each event. This is particularly helpful in reducing the $t\bar{t}h$ background. Applying a Higgs mass window cut of $112.5<m_{bb}<137.5$ GeV, to the invariant mass of b-jets results in a large reduction in the $Zh$ background due to exclusion of the $Z$-peak region. 

\begin{figure}
\begin{tabular}{cc}
		\includegraphics[width=0.49\textwidth]{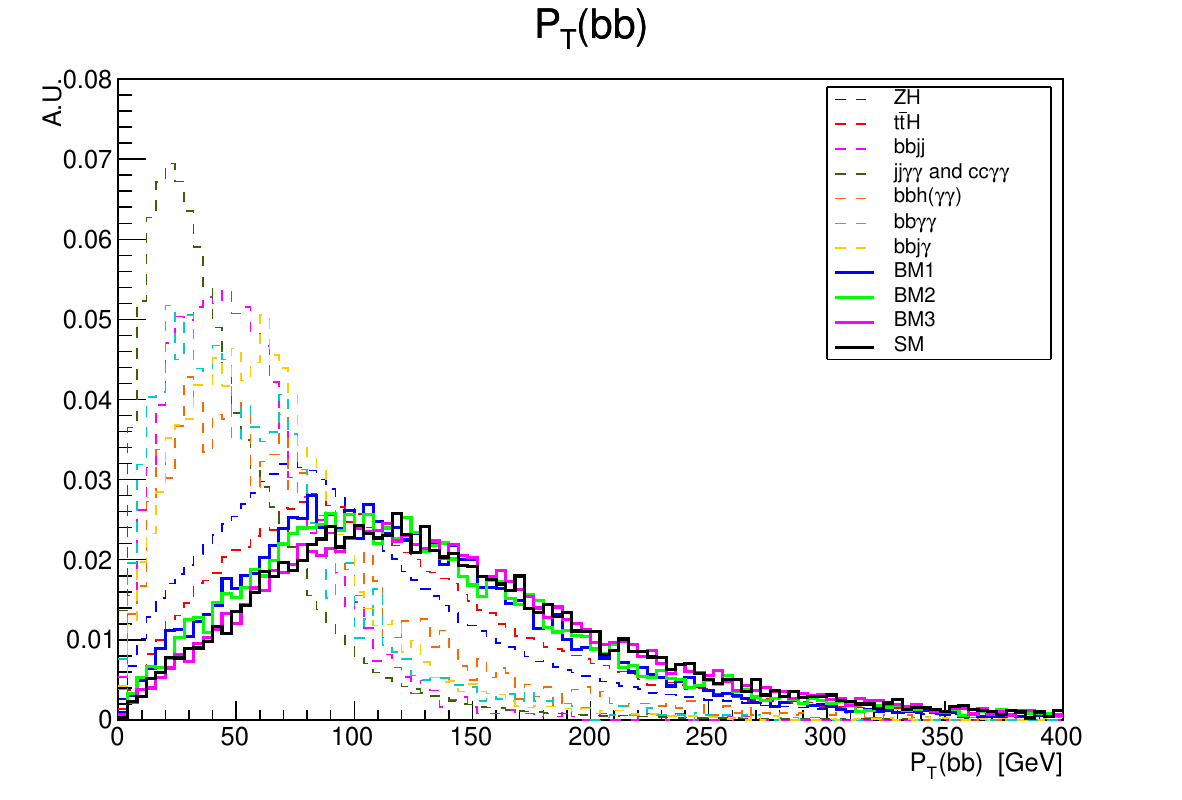}
		\includegraphics[width=0.49\textwidth]{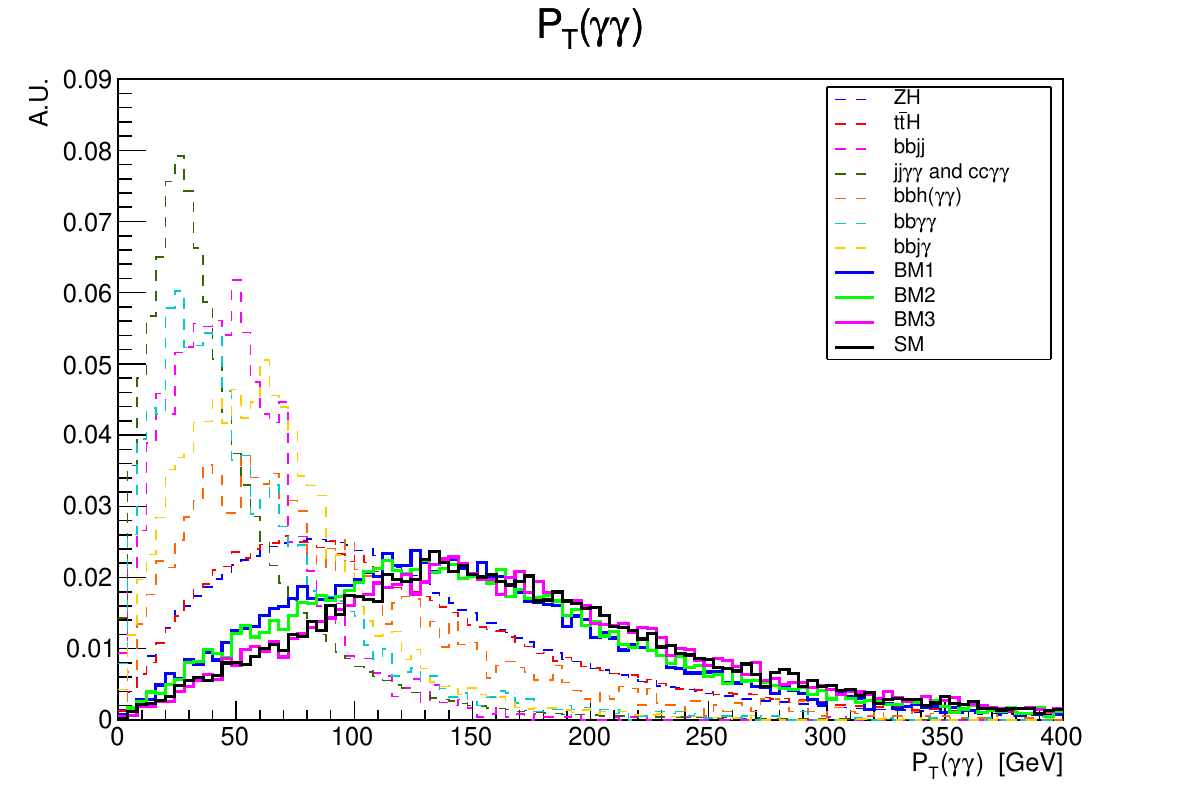}
\end{tabular}
	\caption{Normalized distributions for b-jet-pair and di-photon $p_T$ for signals and various backgrounds as described in the legend. Black solid histogram corresponds to the SM distribution for di-Higgs production. Remaining solid histograms correspond to the three signal benchmarks (BMs) considered. Dashed histograms correspond to various SM backgrounds as indicated in the legend.
	}\label{fig:pTcut}
\end{figure}

The Higgs mass window cut for the di-photon invariant mass is sharper than that for the invariant mass of $b$-jets and helps to reduce the background in all the channels. Furthermore, from the normalized distributions for $b$-jet-pair $p_T$ and di-photon $p_T$ in Fig.~\ref{fig:pTcut} indicate that the signal is favored for $p_T$ values larger than 150 GeV and 140 GeV respectively. Therefore, we further apply these cuts in order to enhance the statistical significance. The resulting efficiencies and cross sections at each stage due to these cuts in our analysis for leading backgrounds and three benchmark (BM) points for the signal are tabulated in Table~\ref{tab:efficiency}.

\begin{table}
	\centering
	\renewcommand{\arraystretch}{1.2}
	\begin{tabular}{|c|c|c|c||c|c||c|c||c|c||}
		\hline
		\textbf{Channel} & \textbf{Pre-selection} & \multicolumn{2}{c|}{\textbf{Basic Cuts}} & \multicolumn{2}{c|}{\textbf{$110<m_{bb}<140$ GeV}} &
		\multicolumn{2}{c|}{\textbf{$pT_{bb}>150$ GeV}} & \multicolumn{2}{c|}{\textbf{$pT_{\gamma\gamma}>140$ GeV}} \\ 
		& \textbf{$\sigma$ (fb)} & \multicolumn{2}{c|}{\textbf{+ $\#$bjet=2;$\#\gamma$=2}} & \multicolumn{2}{c|}{\textbf{$120<m_{\gamma\gamma}<130$ GeV}} &
		\multicolumn{2}{c|}{\textbf{}} & \multicolumn{2}{c|}{\textbf{}} \\ 
		\cline{3-10}
		&  & \textbf{Efficiency} & \textbf{$\sigma$ (fb)} & \textbf{Efficiency} & \textbf{$\sigma$ (fb)} & \textbf{Efficiency} & \textbf{$\sigma$ (fb)} & \textbf{Efficiency} & \textbf{$\sigma$ (fb)} \\
		\hline
		\textbf{Bckgs}  &  &  &  &  &  &  &  &  &  \\ \hline
		$b\bar{b}\gamma\gamma$ & 50500 & 5.64$\times 10^{-4}$ & 28.5 & 1.54$\times 10^{-5}$  & 0.776 &  4.05$\times 10^{-7}$ & 2.04$\times 10^{-2}$ & 3.89$\times 10^{-7}$ & 1.97$\times 10^{-2}$\\ \hline
		$b\bar{b}j\gamma$ & 8424~\footnote{including fake rate of $j \to \gamma$: 0.012\%. } & 4.98$\times 10^{-3}$ & 42.0 & 3.83$\times 10^{-5}$  & 0.322 &  1.56$\times 10^{-6}$ & 1.31$\times 10^{-2}$ & 1.39$\times 10^{-6}$ & 1.17$\times 10^{-2}$\\ \hline
		$c\bar{c}\gamma\gamma$ & 1454.31~\footnote{including fake rate of $c \to b$: 10\%. } & 7.14$\times 10^{-2}$ & 104.0 & 1.64$\times 10^{-4}$ & 0.238 & 3.63$\times 10^{-6}$ & 5.28$\times 10^{-3}$  & 2.90$\times 10^{-6}$ & 4.22$\times 10^{-3}$\\ \hline
		$b\bar{b}h(\gamma\gamma)$ & 35.26 & 3.67$\times 10^{-3}$ & 0.129  & 4.36$\times 10^{-4}$ & 1.54$\times 10^{-2}$ & 8.72$\times 10^{-5}$ & 3.07$\times 10^{-3}$  & 8.33$\times 10^{-5}$ & 2.94$\times 10^{-3}$\\ \hline
		$jj\gamma\gamma$ & 145.33~\footnote{including fake rate of $j \to b$: 1\%. } & 7.90$\times 10^{-2}$ & 11.5  & 1.89$\times 10^{-4}$ & 2.75$\times 10^{-2}$ & 1.48$\times 10^{-5}$ & 2.15$\times 10^{-3}$  & 1.44$\times 10^{-5}$ & 2.09$\times 10^{-3}$\\ \hline
		$t\bar{t}$h$(\gamma\gamma)$  & 38.27 & 2.24$\times 10^{-3}$ & 8.55$\times 10^{-2}$ & 3.22$\times 10^{-4}$  & 1.23$\times 10^{-2}$ & 5.01$\times 10^{-5}$ & 1.92$\times 10^{-3}$ & 2.71$\times 10^{-5}$ & 1.04$\times 10^{-3}$\\ \hline
		$Zh(\gamma\gamma)$  & 1.36 & 3.21$\times 10^{-2}$ & 4.36$\times 10^{-2}$ & 4.56$\times 10^{-4}$ & 6.21$\times 10^{-4}$ & 1.12$\times 10^{-4}$ & 1.52$\times 10^{-4}$ & 1.09$\times 10^{-4}$ & 1.48$\times 10^{-4}$ \\ \hline
		$b\bar{b}$jj & 84.96~\footnote{including fake rate of $j \to \gamma$: 0.012\%. } & 1.09$\times 10^{-2}$ & 0.927 & 1.08$\times 10^{-4}$ & 9.14$\times 10^{-3}$ & 3.25$\times 10^{-6}$ & 2.76$\times 10^{-4}$ & 1.39$\times 10^{-6}$ & 1.18$\times 10^{-4}$ \\ \hline
		\textbf{Total}  & &  & 187.0 &  & 1.40 &  & 4.64$\times 10^{-2}$ &    & 4.19$\times 10^{-2}$ \\ \hline
		\textbf{Sig. BMs}  &  &  &  &  &  &  &  &  &  \\ \hline
		SM & 4.60 & 3.20$\times 10^{-2}$ & 0.147 & 1.36$\times 10^{-2}$ & 6.25$\times 10^{-2}$ & 7.60$\times 10^{-3}$ & 3.50$\times 10^{-2}$ & 7.25$\times 10^{-3}$  & 3.33$\times 10^{-2}$ \\ \hline
		BM1 
		& 9.920 & 3.16$\times 10^{-2}$ & 0.313 & 1.24$\times 10^{-2}$ & 0.123 & 5.30$\times 10^{-3}$ & 5.26$\times 10^{-2}$ & 5.01$\times 10^{-3}$  & 4.97$\times 10^{-2}$ \\ \hline
		BM2 
		& 9.094 & 3.04$\times 10^{-2}$ & 0.275 & 1.25$\times 10^{-2}$  & 0.113 & 5.96$\times 10^{-3}$ &  5.39$\times 10^{-2}$  & 5.73$\times 10^{-3}$ & 5.18$\times 10^{-2}$ \\ \hline
		BM3 
		& 5.329  & 3.16$\times 10^{-2}$ & 0.168 & 1.39$\times 10^{-2}$ & 0.074 & 7.70$\times 10^{-3}$ & 4.10$\times 10^{-2}$ & 7.34$\times 10^{-3}$  & 3.91$\times 10^{-2}$ \\ \hline

	\end{tabular}
	\caption{Cut-flow table for the analysis we perform. Basic cuts refer to generator level cuts described in Eq.~\ref{eqn:basiccuts}. In the cross sections we have multiplied by the following NLO $k$-factors~\cite{Contino:2016spe}: $k_{zh}=0.87$, $k_{t\bar{t}h}=1.3$, $k_{bbjj}=1.08$, $k_{jj\gamma\gamma}=1.43$. Signal benchmarks in the $(g^{(1)}_{HHH}/v, g^{(2)}_{HHH}v)$ plane are as follows:  BM1=$(0.0225,0)$, BM2=$(-0.032,0.0152)$, and BM3=$(-0.141,0.0152)$ }\label{tab:efficiency}
\end{table}

\begin{figure}
	\begin{center}
		\includegraphics[width=0.48\textwidth]{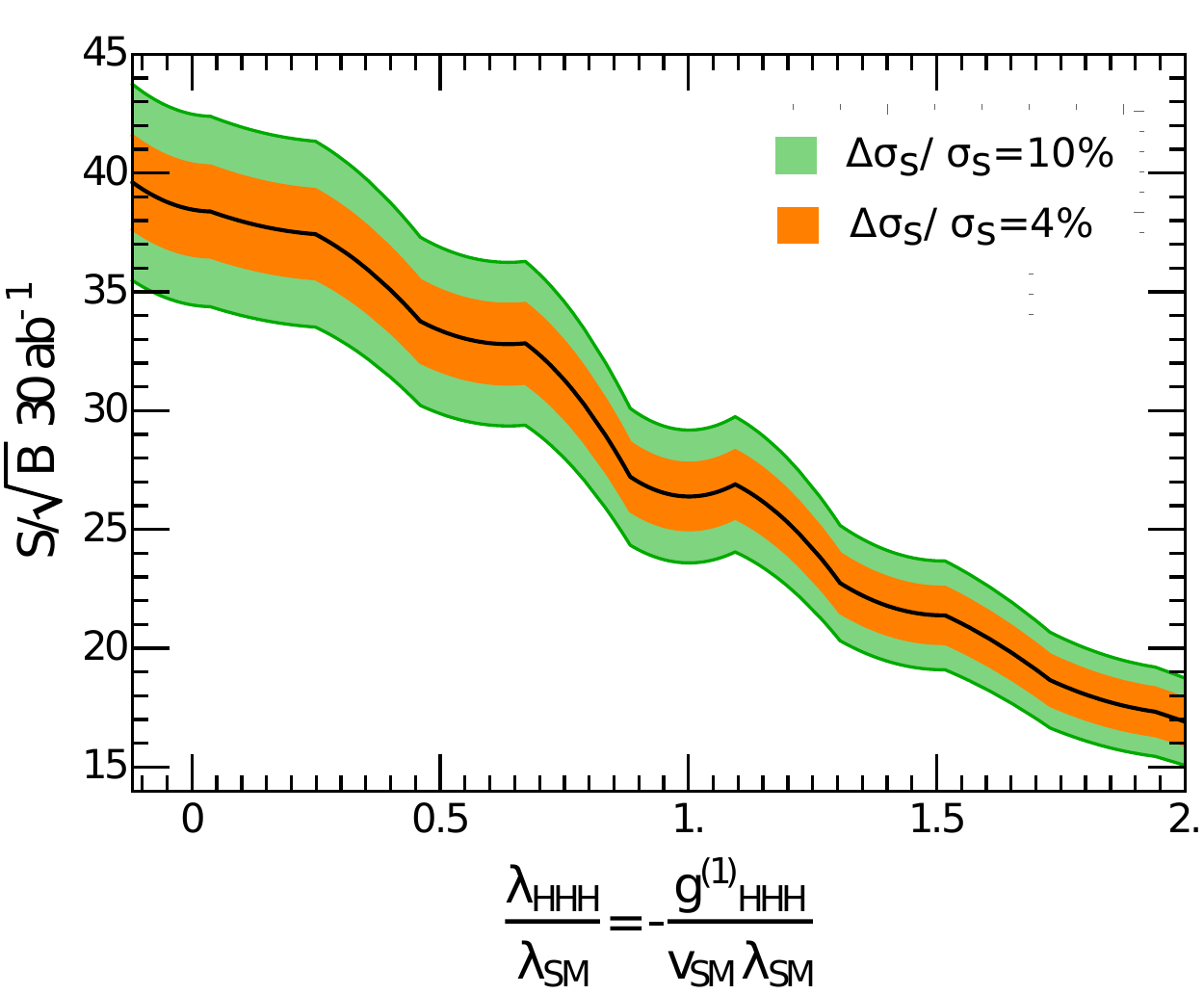}
		\includegraphics[width=0.48\textwidth]{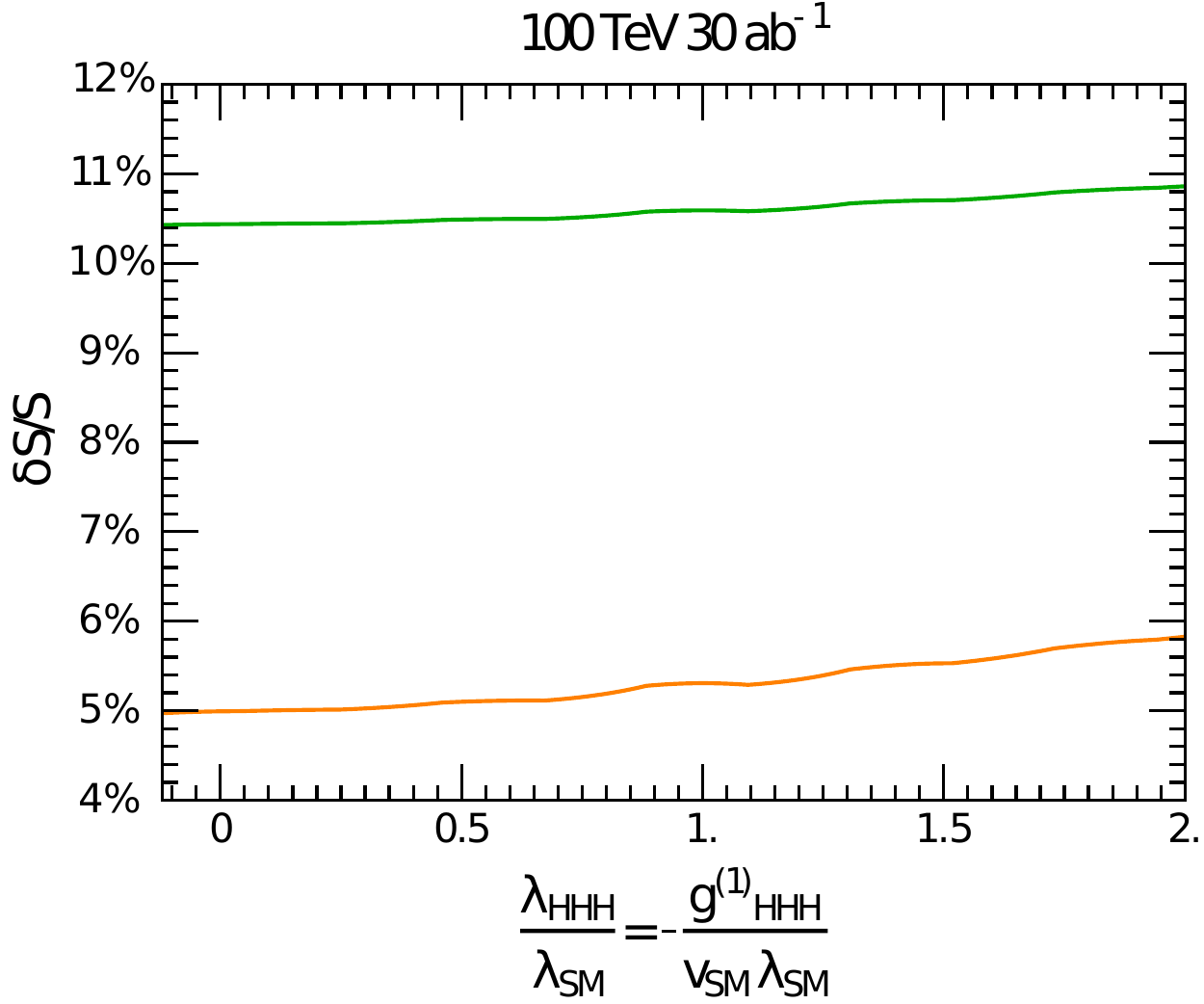}
	\end{center}
	\caption{Left panel: The significance of the di-Higgs process as a function of the trilinear Higgs coupling $\lambda_{HHH} = - g^{(1)}_{HHH}/v$ assuming that the derivative Higgs coupling $g^{(2)}_{HHH}$ is zero. The orange and green bands correspond to the 1$\sigma$ uncertainty in the $S/\sqrt{B}$ with assumptions of the theoretical uncertainty for the di-Higgs production cross-section to be 4\% and 10\% respectively. Right panel: The percentage uncertainties on the measured number of signal events varies with the value of trilinear Higgs coupling. Orange and green lines correspond to theoretical uncertainties of 4\% and 10\% respectively.\label{fig:ratiosig}}
\end{figure}

We first investigate the sensitivity of the trilinear Higgs coupling $\lambda_{HHH} = - g^{(1)}_{HHH}/v$ in the absence of the derivative Higgs coupling $g^{(2)}_{HHH}$. In this case, we recover the scenario widely discussed in the literature: how to probe the deviation of the $\lambda_{HHH}$ from its SM value $\lambda_{SM}$ at the future collider. Compared with the work in~\cite{Barr:2014sga}, we obtain comparable significance of about $8.25\sigma$ for the SM di-Higgs production for luminosity of 3 ab$^{-1}$. This corresponds to the significance of $\sim 26\sigma$ for 30 ab$^{-1}$ as can be seen from the black line in the left panel of Fig.~\ref{fig:ratiosig}, where we plot the $S/\sqrt{B}$ for $30\ {\rm ab}^{-1}$ and zero derivative interaction.

We also estimate the uncertainty in the value of $S/\sqrt{B}$ by taking into account the statistical uncertainty for the signal and background as well as the theoretical uncertainty on the di-Higgs production cross-section. It turns out that for a $30\ {\rm ab}^{-1}$ luminosity, the statistical uncertainty in the number of signal events due to Poisson fluctuations is around 3\%, which is less than the 10\% theoretical uncertainty coming from the infinite top mass approximation, the scale, and the PDF uncertainties~\cite{Contino:2016spe}. The 1$\sigma$ uncertainty due to this is denoted by the green band in the left panel of Fig.~\ref{fig:ratiosig}. However, the latest estimation on the theoretical uncertainties places them as low as 4\%~\cite{slides}. Therefore, we also include this case denoted by orange band in the plot shown in the left panel of Fig.~\ref{fig:ratiosig}.
	
The right panel of Fig.~\ref{fig:ratiosig} represents the percentage uncertainties for the measured number of signal events as a function of the ratio of the triple Higgs coupling to its SM predicted value. Orange and green lines here correspond to the theoretical uncertainty of 4\% and 10\% respectively. As expected from the above quoted numbers, the theoretical uncertainty dominates except where the ratio of triple Higgs couplings is close to 2.5, where the cross section for di-Higgs production is the lowest leading to enhanced uncertainty due to Poisson fluctuations.
%slides https://indico.cern.ch/event/618254/contributions/2833246/attachments/1583347/2502527/HH_2ndFCCWS_.pdf

Here we comment on the validity of EFT in our collider analysis. The EFT breaks down when the parton collision center of mass energy approaches the scale of the cutoff scale $M=2$ TeV. Therefore, we should in principle add a cut on the kinematic variables like invariant mass of di-Higgs to only keep the events produced in low energy regime to make our EFT analysis valid. The di-Higgs spectrum is peaked at an invariant mass $m_{hh}$ near the two higgs threshold indicating our EFT approach should be valid (i.e. the processes considered have energy well below our cutoff of 2 TeV). Additionally we have investigated the number of events below 1 TeV, 1.5 TeV, and 2 TeV for three benchmark points: the SM, $\lambda_{HHH}/\lambda_{SM}=2$, $\lambda_{HHH}/\lambda_{SM}=-0.9$, and finding the results in Table.~\ref{tab:mhhcut}. As there are only a small number of outlying events with higher energies these numbers support the assertion that the EFT approach is valid in our Monte Carlo simulation. One should note that even if the heavy particles were to be discovered at higher energies that in order to extract the trilinear couplings of the SM Higgs one would still employ an EFT. Such a procedure is analogous to the use of an effective four fermion theory for flavor physics where the heavy $W$s have been integrated out of the theory in favor of unrenormalizable operators.
\begin{table}
\begin{tabular}{| c | c | c | c |}
    \hline
    $\frac{\lambda_{HHH}}{\lambda_{SM}}$ & $m_{hh}>1 $TeV & $m_{hh}>1.5$ TeV & $m_{hh}>2 $TeV\\
    \hline
    1 & 2.5\% & 0.38\% & 0.16\% \\
    \hline
	2 & 5.1\% & 1.0\% & 0.35\% \\
	\hline
    -0.9 & 1.3\% & 0.26\% & 0.05\% \\
    \hline
\end{tabular}
	\caption{The percentage of events with $m_{hh}$ above 1, 1.5 and 2 TeV.}\label{tab:mhhcut}
\end{table}

%\textcolor{red}{We have simulated a smaller data set with the addition of this cut to select the events with $m_{hh} <1.5$, and we found that this cut only alters the efficiency for the signal events by less than 0.5\%. Thus our analysis is a good approximation for the kinematic region that is valid for the EFT. One should note that even if the heavy particle were to be discovered at higher energies, to extract the trilinear couplings of the SM Higgs one would formally still employ an EFT in order to avoid the introduction of unphysical large logarithms. (TC: I changed the language a bit here because im confused as to what you're saying. Did you go back and redo some simulations to test the effect of the hard cut on energy? Or did you redo the full analysis. The language seemed to imply the former so I've rewritten things to make the process clearer, but since I don't know exactly what you did this may be wrong)}

\subsection{Determination of Wilson Coefficients}

 \begin{figure}[h]
 	\begin{center}
 		\includegraphics[width=0.49\textwidth]{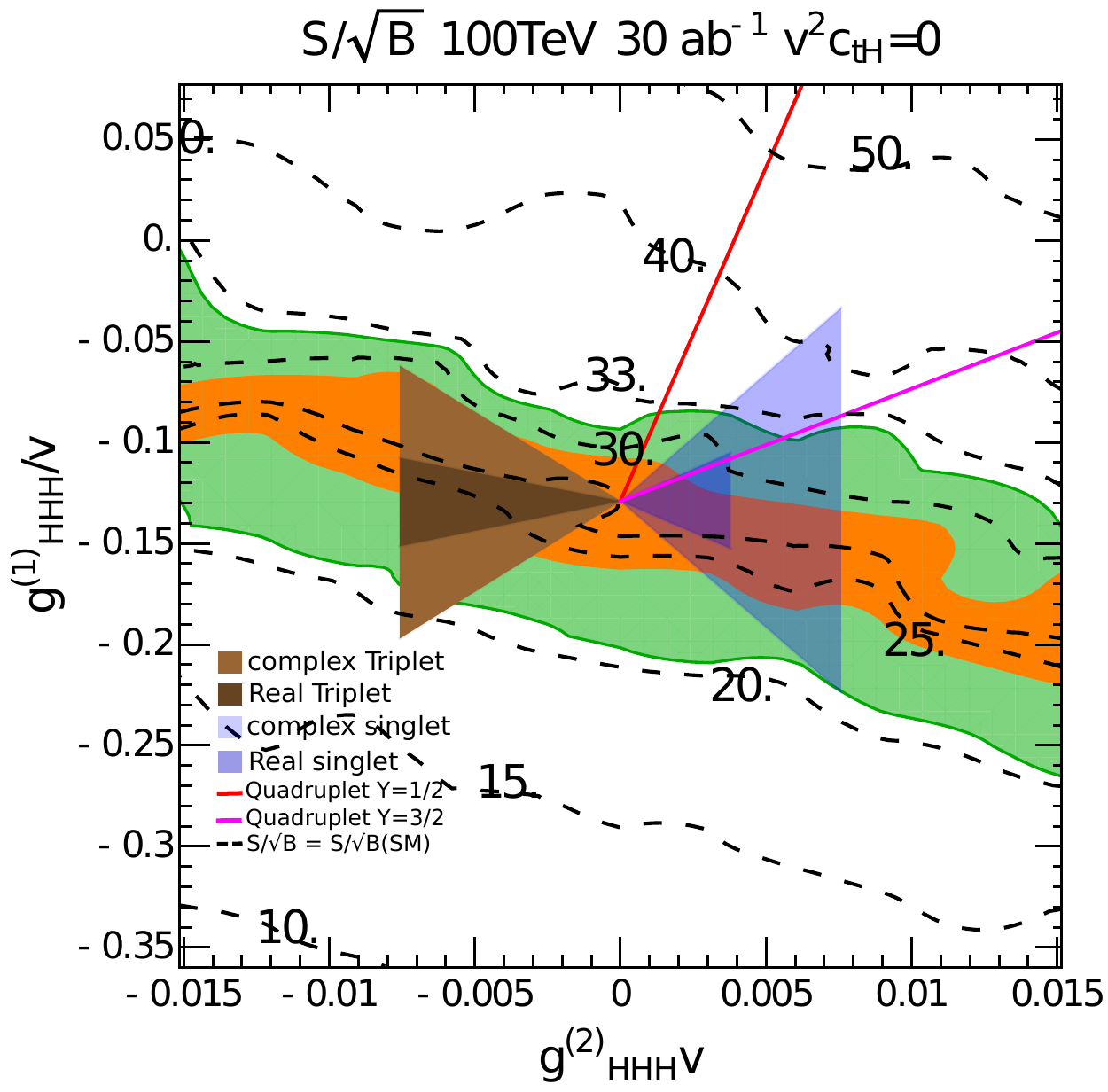}
 		\includegraphics[width=0.49\textwidth]{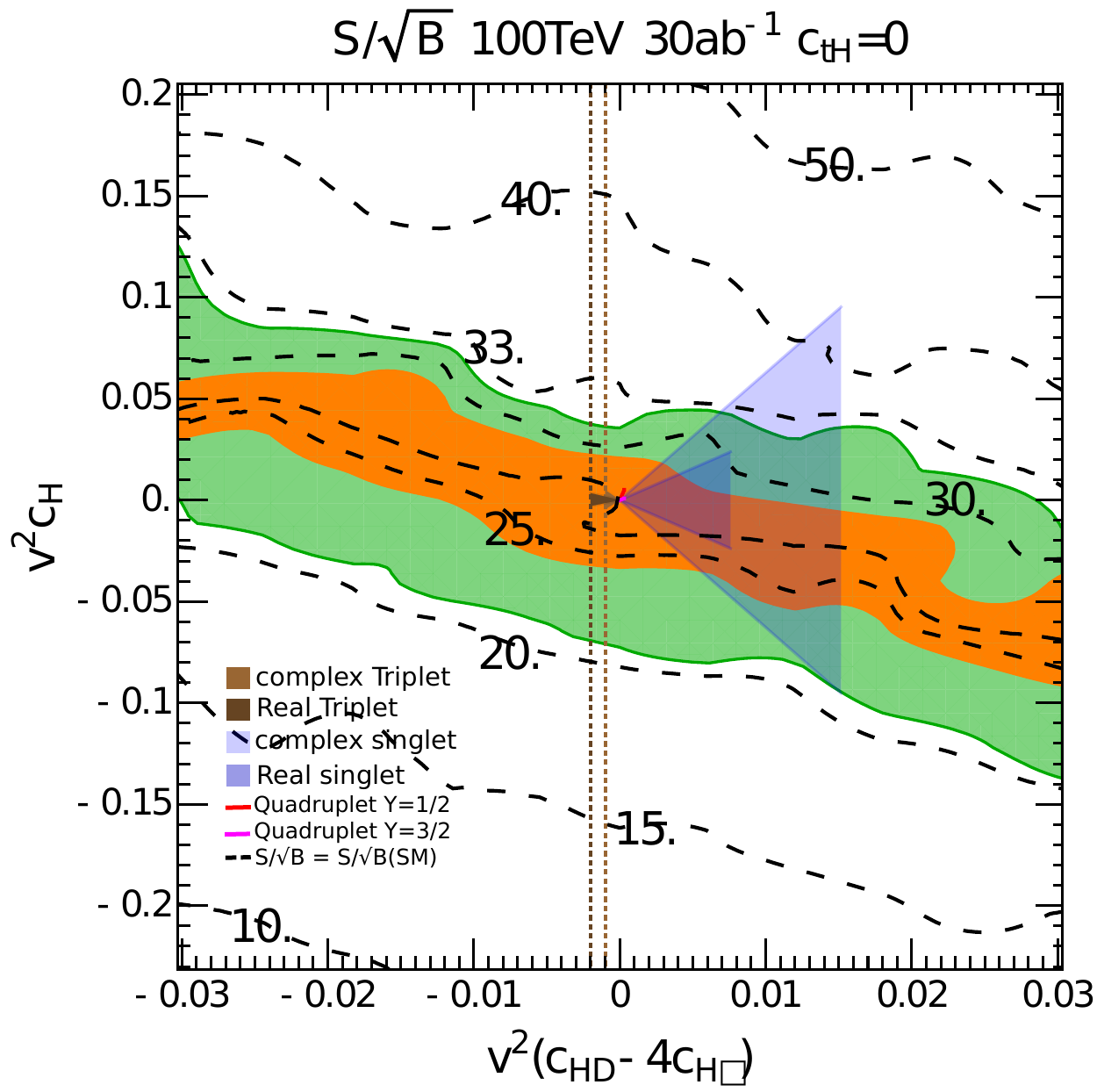} 
 	\end{center}
	 \caption{Black dashed contours denote statistical significance, $S/\sqrt{B}$, for identifying the signal at 100 TeV with integrated luminosity of 30~${\rm ab}^{-1}$. Left panel: The significance contours are plotted in the $g_{HHH}^{(1)}/v$ vs. $g_{HHH}^{(2)}$ plane, the shaded region is constrained by dimensionless couplings in the Lagrangian within the range $\pm 4\pi$ for couplings with mass dimension within the range $\pm 1$ TeV and cutoff scale $M=2$ TeV. The light and dark shaded brown and blue regions are allowed by all the global fit constraints. The Red line and magenta line corresponds to quadruplet model with $Y=$ 1/2 and 1/3 respectively. Orange and green regions correspond to the 1$\sigma$ uncertainty on the significance with assumptions of the theoretical uncertainty for the di-Higgs production cross-section to be 4\% and 10\% respectively. Right panel: The significance contours are plotted in the $v^2c_{H}$ vs. $v^2(c_{HD}-4c_{H\square})$ plane. The darker brown and light brown dotted lines on the right panel correspond to the Wilson coefficient constraints from the Higgs coupling measurements and the $T$-parameter in the real and complex triplet models. Shaded regions on the right have the same meaning as in the left panel. Both plots are with $c_{tH}=0$ and the SM limit in both is located at $(0,0)$ with $S/\sqrt{B} \sim 26 $.}\label{fig:sig3}
 \end{figure}

 \begin{figure}[h]
 	\begin{center}
 		\includegraphics[width=0.48\textwidth]{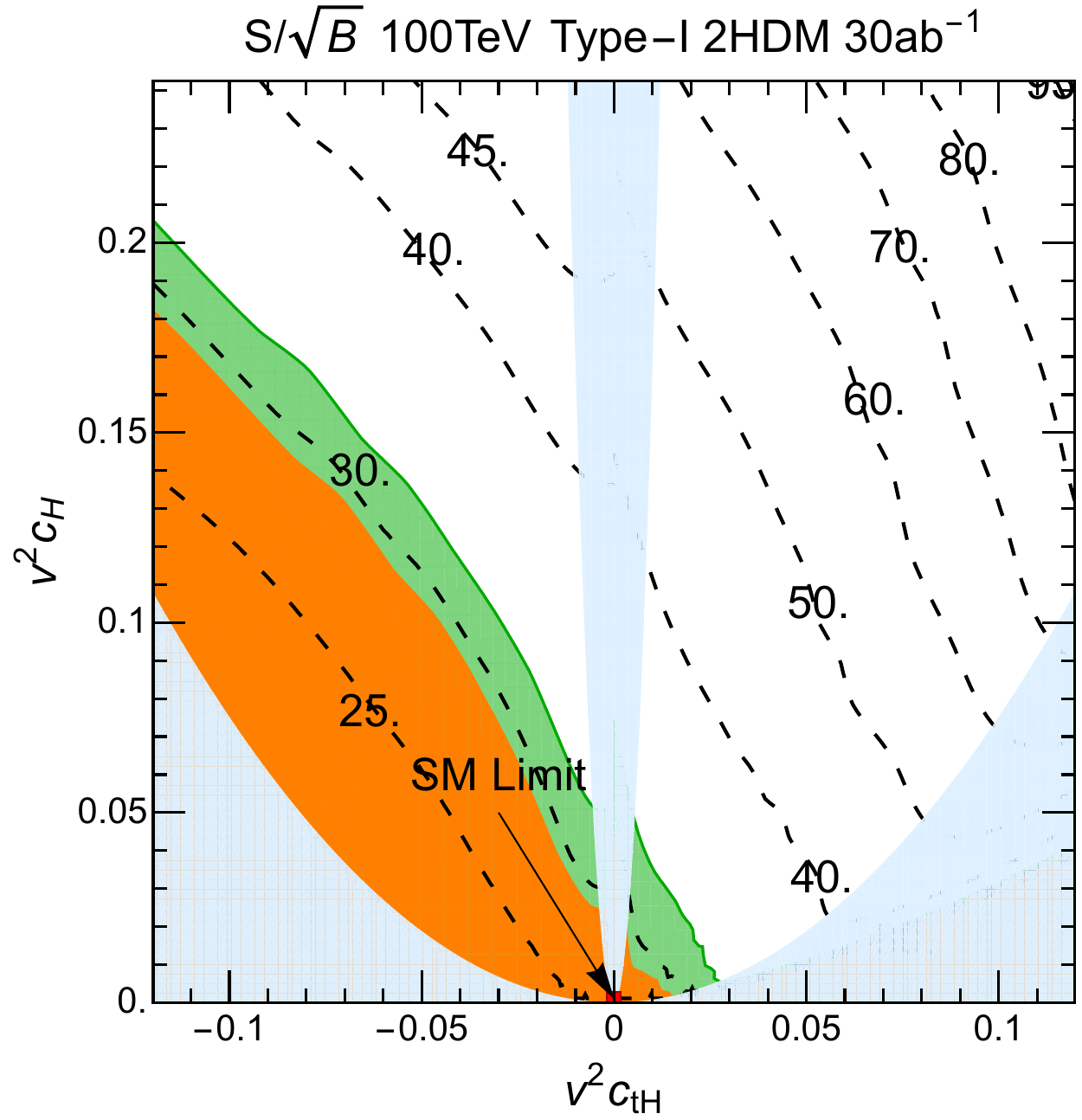}
 		\includegraphics[width=0.48\textwidth]{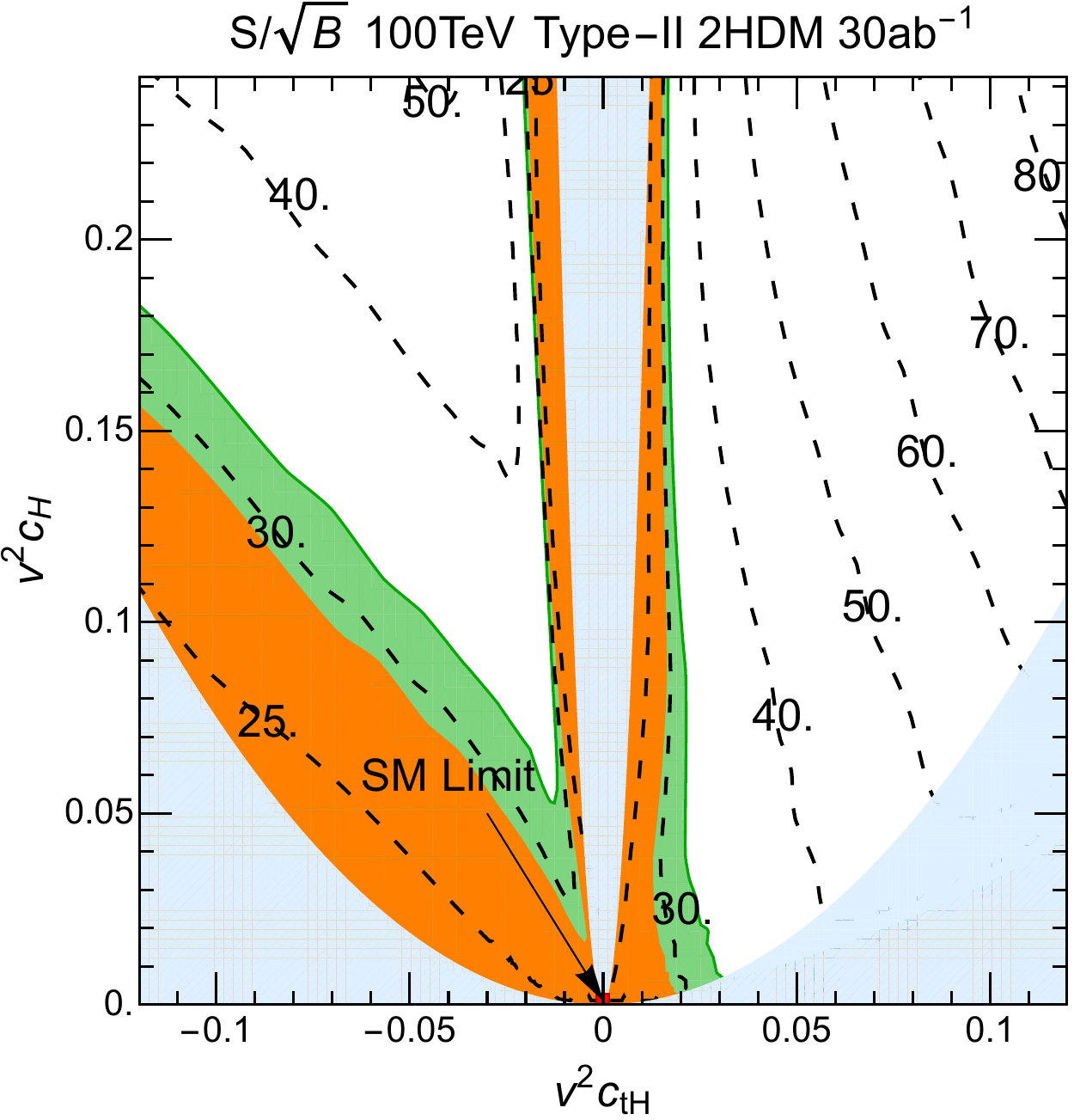}
 	\end{center}
 	\caption{Dark cyan dashed contours denote statistical significance, $S/\sqrt{B}$, for identifying the signal at 100 TeV with integrated luminosity of 30~${\rm ab}^{-1}$. The left and right plots represents Type-I and Type-II 2HDM respectively.
	The light blue regions correspond to the parameter regions in $\tan\beta$ which has been ruled out by experimental data from flavor physics. The orange and green regions are within the SM 2$\sigma$ uncertainty with assumption of the percentage uncertainty of di-Higgs production cross section equal to 4\% and 10\% respectively. 
 		\label{fig:sig2hdm}}
 \end{figure}

Equation~\ref{eq:lorentzcoeff} and Table~\ref{tab:EFTtable} demonstrate it is necessary to investigate the discovery potential at the 100 TeV collider when both the deviation of the $\lambda_{HHH} $ coupling from the SM value, and non-zero $g^{(2)}_{HHH}$ exist. Turning on the derivative Higgs coupling $g^{(2)}_{HHH}$ will change the significance of the di-Higgs signatures. In Fig.~\ref{fig:sig3} we present the reach of the 100 TeV collider with integrated luminosity of 30 ${\rm ab}^{-1}$ in the space of $g^{(1)}_{HHH}-g^{(2)}_{HHH}$ in the left panel as well as in the space of Wilson coefficients $c_H$ and $c_{HD}-4c_{H\Box}$ in the right panel, each with $c_{tH}=0$. 
The left and right panels of the Fig.~\ref{fig:sig3} are not independent. Their values are connected by Eq.~\ref{eq:lorentzcoeff}, where the contours in the right panel are essentially rotated around the SM values as governed by the Eq.~\ref{eq:lorentzcoeff}. This represents only a class of models, in which
$c_{tH}$ is not important, for example, singlet, triplet and quadruplet models. We plot the statistical significance contours for 2HDMs in $c_{tH}-c_{H}$ space as shown in separate plots of Fig.~\ref{fig:sig2hdm}. $c_{tH}=0$ corresponds to $\tan\beta\to\infty$, which is outside the experimental bounds on $\tan\beta$ in 2HDMs.

Fig.~\ref{fig:sig3} shows the allowed parameter regions in singlet, triplet and quadruplet models, which overlap within the significance contours. 
In these models, according to Table~\ref{tab:EFTtable}, the Wilson coefficients $c_H$ and $c_{HD}-4\,c_{H\Box}$ are not independent. 
More specifically, they are related by linear relations such as $c_H \simeq \lambda_{HS(\Phi)}(c_{HD}-4\,c_{H\Box})$. This linear relation then implies that the boundaries of these regions are governed by the input perturbative limit $|\lambda_{HS(\Phi)}|\le4\pi$ and are straight lines as can be seen in Figure~\ref{fig:sig3}.
The values of the dimensionless Higgs scalar couplings, such as $\lambda_{HS}$, $\lambda_{H\Phi}$, determine the 
slopes of the parameter region in each model. 
For example, in the real singlet case, along the boundary of the parameter region, the Higgs scalar coupling $\lambda_{HS}$ should be around $\pm 4\pi$.
In the region far from the boundary, the dimensionless Higgs scalar couplings appearing in $c_H$ should be small. 
We choose $c_{tH}$ to be equal to zero in these two plots.
This condition is automatically satisfied by singlet and quadruplet models, and also approximately satisfied by triplet models.
This is because $c_{tH}$ in triplet models is suppressed by the coupling $g^2$ which is constrained to be very small by EWPD due to its relation to the $T$-parameter.

In addition to the allowed region in each model, we also illustrate the region that will generate the expected significance within the 2$\sigma$ uncertainties around SM value. In principle one should derive the prospective confidence level contour for the parameter space that are consistent with SM prediction in the future experiments. However, this requires the scanning of a fine grid to obtain the selection efficiency of each point, which is beyond the scope of our study. %our computation power. 
Therefore we simply estimate that this 2$\sigma$ region roughly gives the region that is hard to differentiate from the SM in the future experiments.
%{\color{green} In addition to the allowed region in each model, we also estimate the 2$\sigma$ uncertainties around Standard model points on the 2D parameter space. 
%The full statistical analysis on discovering the 2D parameter space could be very complicated, which depends on the model parameter choices. 
%Since we note that the cross section in UV models is not sensitive to the coupling $g^{(2)}_{HHH}$, we could neglect 
%the $g^{(2)}_{HHH}$ dependence in our estimation procedure. 
%Furthermore, we could take an rough estimation that if the predicted NP cross section  lies out of the uncertainties of the SM prediction, it is likely to come from the NP contribution. 
%To have an estimation on whether the involves in complicity from We take the stre with the assumption of the SM model value.} 

One can observe that, the future di-Higgs experiment is not sensitive to the $c_{H\square}$ and $c_{HD}$ which have already been strongly constraint by the EWPD. On the other hand, it can constrain the value of $c_{H}$. Depending on the theoretical uncertainties that can be achieved, it may also be possible to exclude some parameter space of the singlet models, which represents the region outside the 2$\sigma$ region.

The case of the 2HDM is much more promising for distinguishing between the SM and the NP model as $c_{tH}$ is non-zero. We demonstrate the significance for 100 TeV collider at 30 ${\rm ab^{-1}}$ integrated luminosity in $v^2 c_{tH}$ vs $v^2 c_{H}$ plane, shown in Fig.~\ref{fig:sig2hdm}. Both $v^2 c_{tH}$ vs $v^2 c_{H}$ depend on $\tan\beta$. Here  we choose the range of $\tan\beta$ such that it satisfies the constraints from flavor physics according to Ref.~\cite{Enomoto:2015wbn}. 
This rules out some parameter regions as shown in Fig.~\ref{fig:sig2hdm} by blue regions. We note that the significance in the 2HDM is generally larger than that of the singlet, triplet and quadruplet models due to typical enhancement from the Yukawa couplings, and it is very likely to observe a significant deviation from the SM signal. We also find that, unlike the singlet and triplet, signal significances in the 2HDM are much more enhanced compared to the ones in the SM.
The plots also show that the contours of significance of two types of 2HDMs are different despite the coupling to up-type quarks being the same in both Type I and Type II, the reason being that we are using the $bb\gamma\gamma$ final state and the branching ratio of $h\to bb$ are different between the two versions of the 2HDM.

From Fig.~\ref{fig:sig3} and Fig.~\ref{fig:sig2hdm}, if we limit ourselves in these models with all the heavy particles integrated out, the di-Higgs process puts additional constraints on the scalar model parameters. Our analysis in Fig.~\ref{fig:sig3} and Fig~\ref{fig:sig2hdm} shows that the Complex singlet and 2HDM (triplet and quadruplet) scalar models are the most (least) sensitive, among those resulting from the models under consideration, to the collider search. As a consequence, the di-Higgs process probes the allowed region of $c_H$, and thus the Higgs scalar couplings in the UV models.

\subsection{Exploring Parameter Region in UV Models}

 \begin{figure}
 	\begin{center}
 		\includegraphics[width=0.49\textwidth]{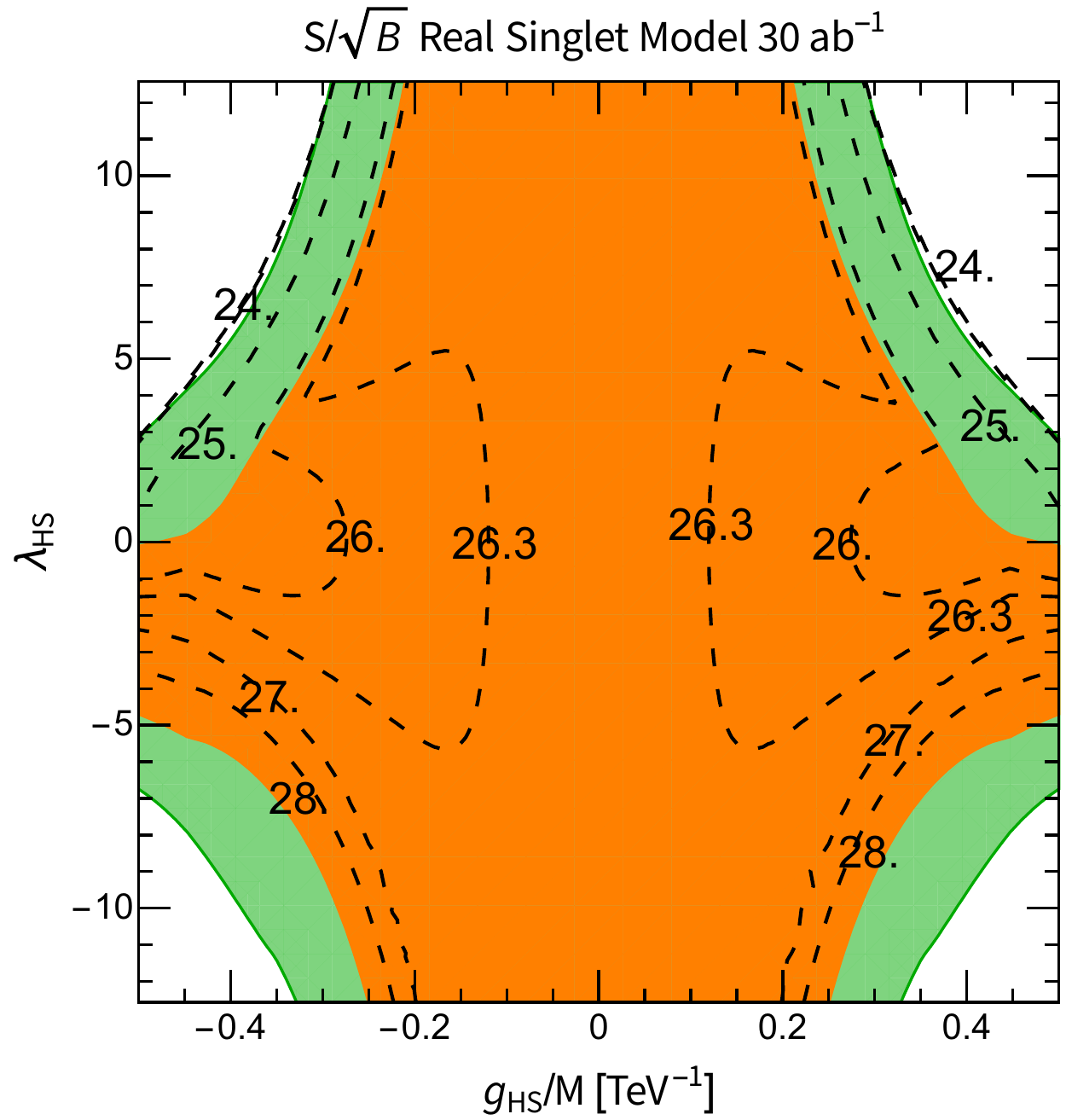}
 		\includegraphics[width=0.49\textwidth]{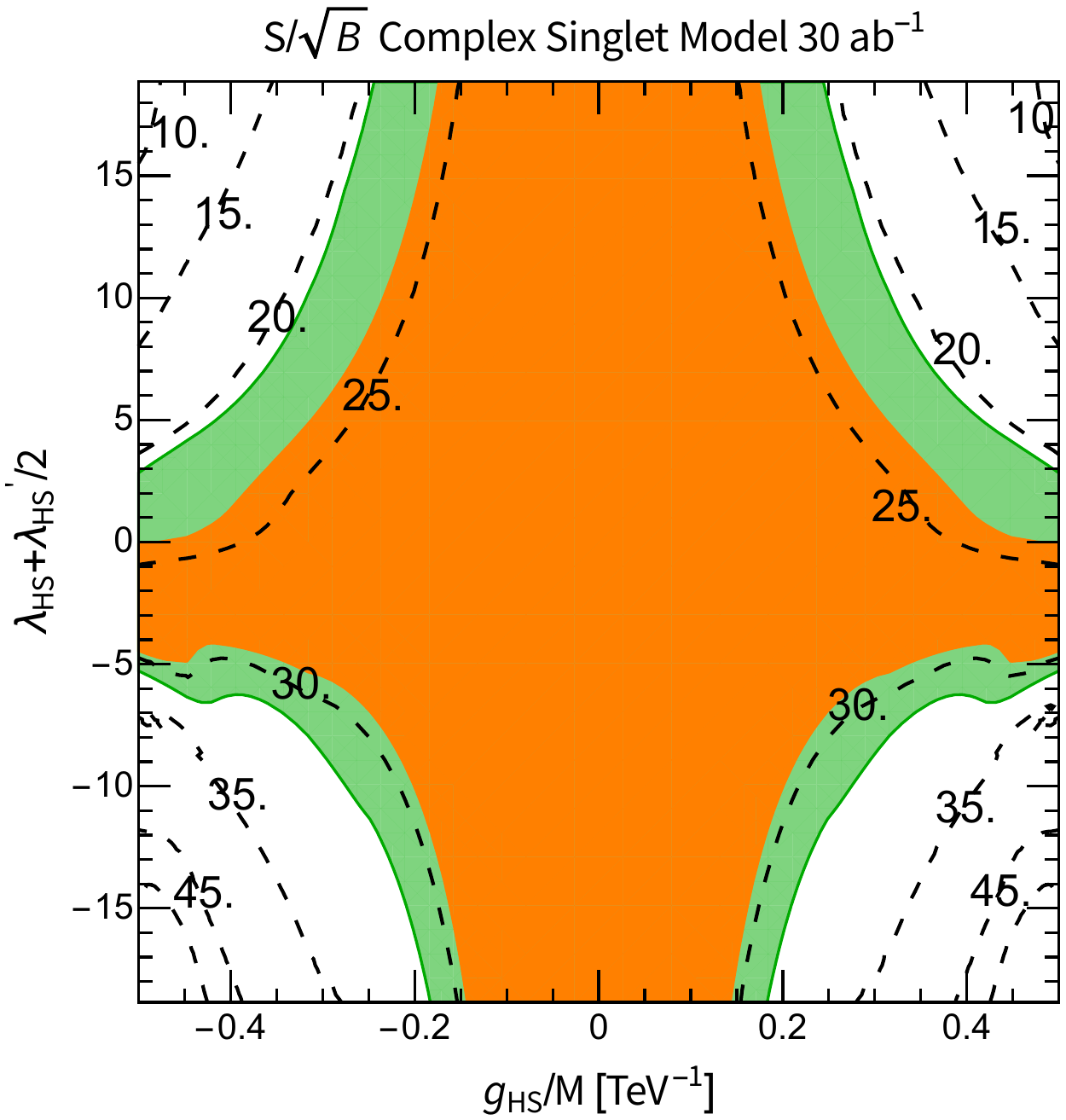}
 	\end{center}
 	\caption{The discovery potential of the model parameters $(g_{HS}, \lambda_{HS})$ in the real (left panel) and complex (right panel) singlet models. The contours correspond to the significance given integrated luminosity of 30 ~${\rm ab}^{-1}$. The orange and green regions are with in the SM 2$\sigma$ uncertainty with assumption of the percentage uncertainty of di-Higgs production cross-section equal to 4\% and 10\% respectively. 
 		\label{fig:modelbounds}}
 \end{figure}
 
 We project the sensitivity of the Wilson coefficients into the parameter space corresponding to the models under consideration. In the real singlet model, the parameter space of the effective coefficients allowed is indicated by the light blue region in Fig.~\ref{fig:sig3}, can be probed with $S/\sqrt{B}$ more than $25$, while in the complex singlet model, the Wilson coefficients resulting from integrating out the complex singlet can be probed to $S/\sqrt{B}$ values higher than even $40$. In Fig.~\ref{fig:modelbounds}, we show the possible reach of the model parameters $(\lambda_{HS}, g_{HS})$ in the real singlet model, and $(\lambda_{HS}+\lambda'_{HS}/2, g_{HS})$ in the complex singlet model, given $30$ ab$^{-1}$ luminosity data set. One can see that, most of the region in the singlet and triplet models are within the 1$\sigma$ uncertainty band for $S/\sqrt{B}$ reach for the SM, so that they are hard to differentiate from the SM.
%The full parameter regions of real singlet model and complex singlet model, in which all the heavy particles are assumed to be integrated out, can be explored respectively given $3$ ab$^{-1}$ luminosity data set.

\begin{figure}
	\begin{center}
		\includegraphics[width=0.49\textwidth]{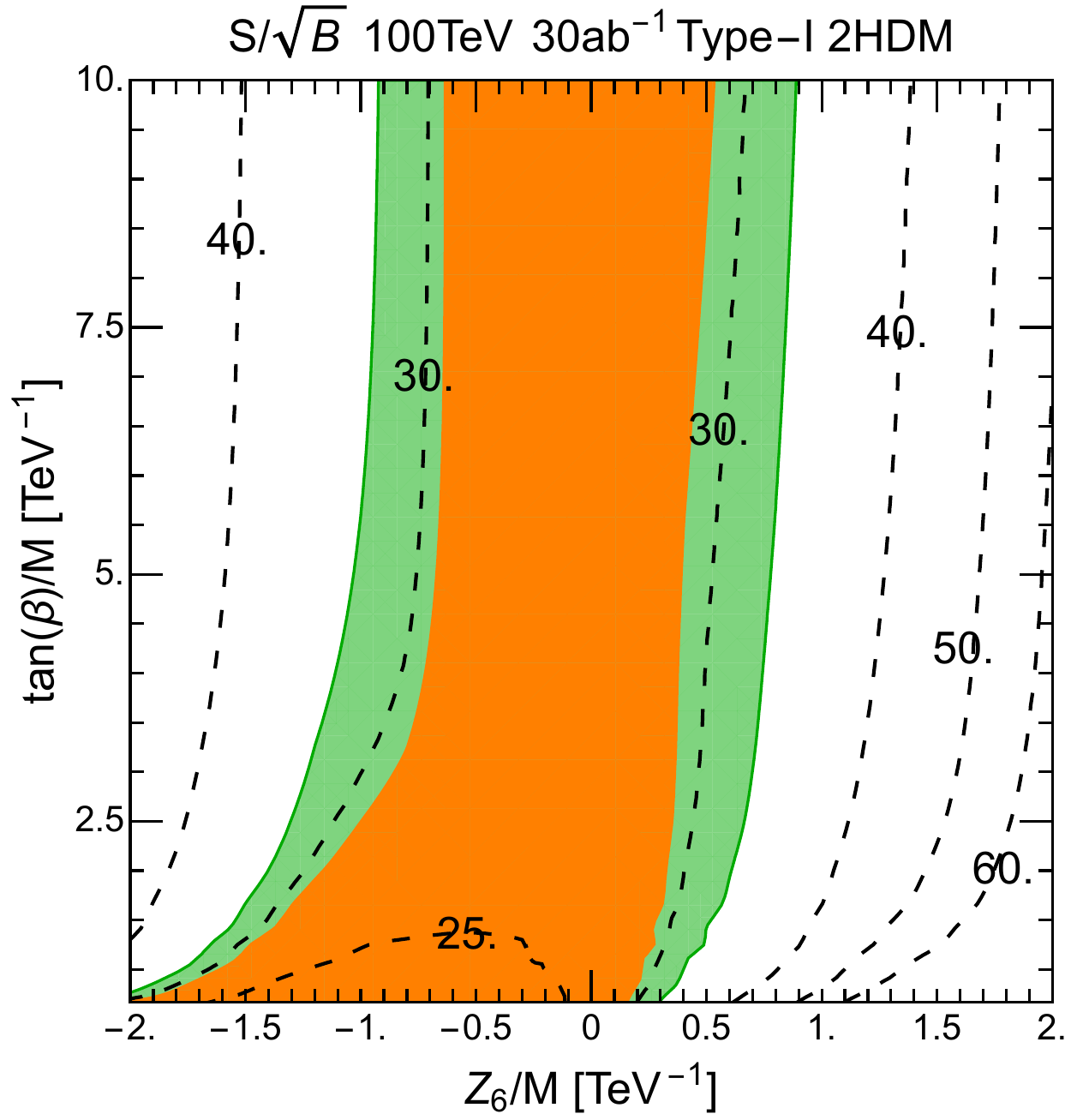}
		\includegraphics[width=0.49\textwidth]{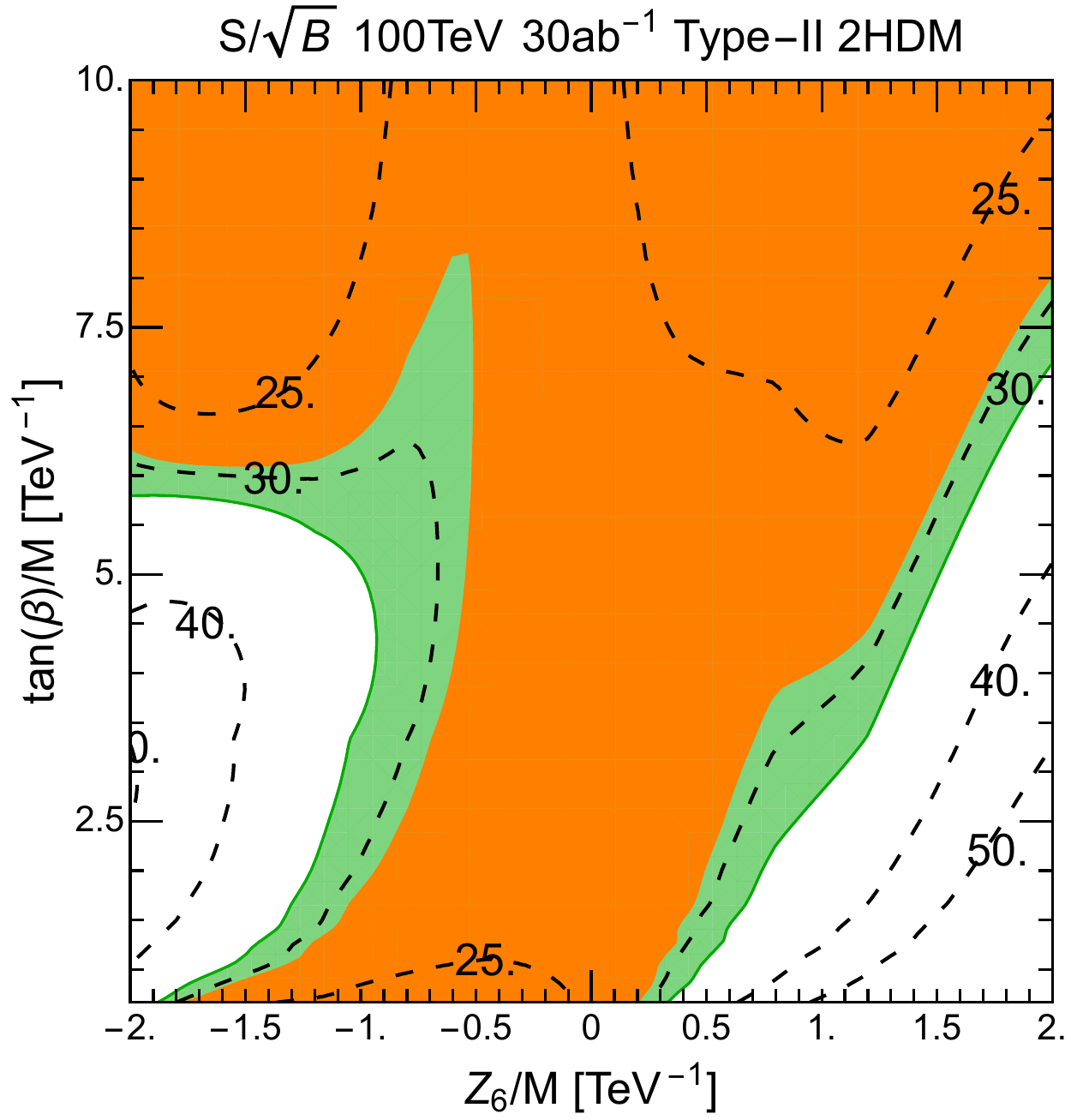}
	\end{center}
	\caption{The discovery potential of the model parameters $(Z_6, \tan\beta)$ in the Type-I (left panel), Type-II (right panel) 2HDM. The contours correspond to the significance given integrated luminosity of 30~${\rm ab}^{-1}$. The orange and green regions are with in the SM 2$\sigma$ uncertainty with assumption of the percentage uncertainty of di-Higgs production cross-section equal to 4\% and 10\% respectively. 
		\label{fig:modelbounds2hdm}}
\end{figure}

The 2HDM, owing to its preservation of custodial symmetry, resides on the line $c_{HD}=4c_{H\Box}=0$ (up to the assumptions made in this paper, that is a tree-level dimension-six analysis). Therefore, the Higgs coupling measurements and the electroweak precision tests do not place strong constraints on the model parameters. On the other hand, the di-Higgs signature starts to provide a strong constraint on $c_H$. 
%
%Similar to the singlet models, the 2HDM parameters can be probed with a statistical significance of  $10\sigma$. 
%
In Fig.~\ref{fig:modelbounds2hdm} we show the significance contour on the model parameter $Z_6$ vs $\tan\beta$ plane  for Type-I model and  Type-II model with the $30$~$ab^{-1}$ luminosity.
Note that when $Z_6 = 0$, the SM limit is recovered (see Table~\ref{tab:EFTtable}).
We also find that 
%with $3$~ab$^{-1}$ luminosity, all the parameter space of interest can be explored for Type-I model and Type-II model. 
in the Type-II model, for negative $Z_6$ and large $tan\beta$ (left top corner in the right plot in Fig.~\ref{fig:modelbounds2hdm}), the significance approaches to the SM value. 
%This is because at large $tan\beta$  the Higgs to b quark coupling is increased, and it reduces the Higgs to b decay branching ratio, which ameliorate the increasing di-Higgs production rate. 
This is because the decreasing of the Higgs to b quark coupling reduces the Higgs to b decay branching ratio, which ameliorate the increasing of the di-Higgs production rate.

In the real and complex triplet models, both $c_{HD}$ and $c_{H\Box}$ in the EFTs obtained by integrating out real and complex triplet models are very tightly constrained as shown by the vertical dashed lines, shown in Fig.~\ref{fig:sig3} (right panel). These vertical darker and lighter green lines represent the $3\sigma$ bounds allowed by the Higgs data global fit on the Wilson coefficient linear combination of $c_{HD}-4c_{H\Box}$ for the real and complex triplet model respectively. The reason that these stringent bounds only exist for the triplets and not the singlets is that the coefficient $c_{HD}$ is connected with custodial symmetry breaking and is tightly constrained by the electroweak precision parameter $T$. 

As Table~\ref{tab:EFTtable} denotes, the $c_{HD}$ and $c_{H\Box}$ are tightly related for the triplet models and therefore the stringent bounds on $c_{HD}$ translate into stringent bounds on the $c_{HD}-4c_{H\Box}$ as well. In the case of the singlet models, there are no couplings of the singlets to the gauge bosons resulting in $c_{HD}$ being identically zero as indicated in Table~\ref{tab:EFTtable}, liberating them from these constraints suffered by the triplet models. 
As a result of these, $c_H$ is also strongly constrained from the small allowed values of $c_{HD}-4c_{H\Box}$, as shown in Fig.~\ref{fig:sig3} (right panel). 
However, the dimensionless Higgs potential parameters, such as $\lambda_{H\Phi}$ and $\lambda$, are still very loosely constrained due to $c_H \sim \frac{g^2}{M^4}\lambda_{H\Phi}$. 
Therefore, it is very hard for us to extract the Higgs scalar couplings from the $c_H$ operator, because the deviation of the Higgs coupling from the SM value is very small in the triplet case. 

\begin{figure}
	\begin{center}
		\includegraphics[width=0.48\textwidth]{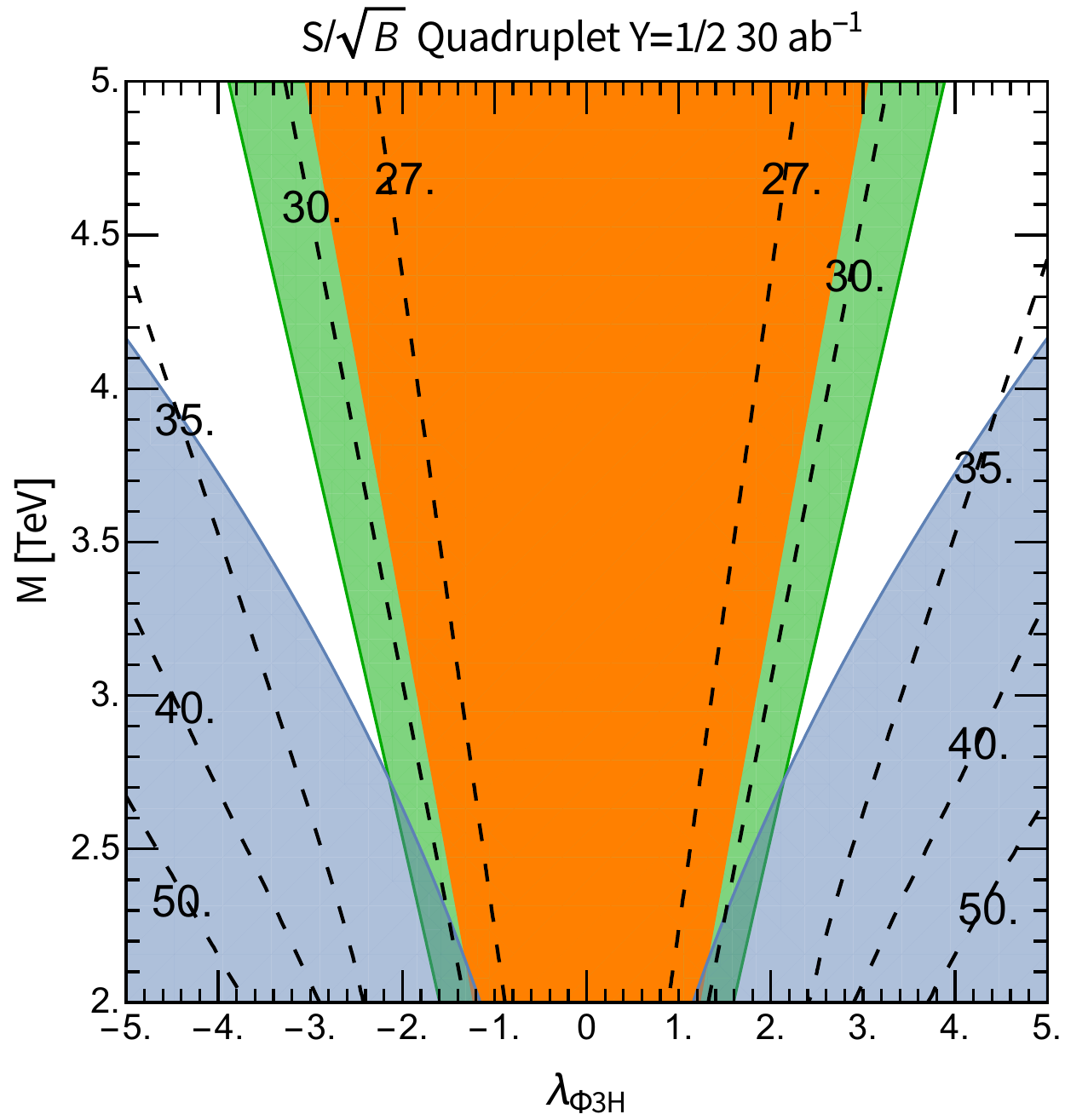}
		\includegraphics[width=0.48\textwidth]{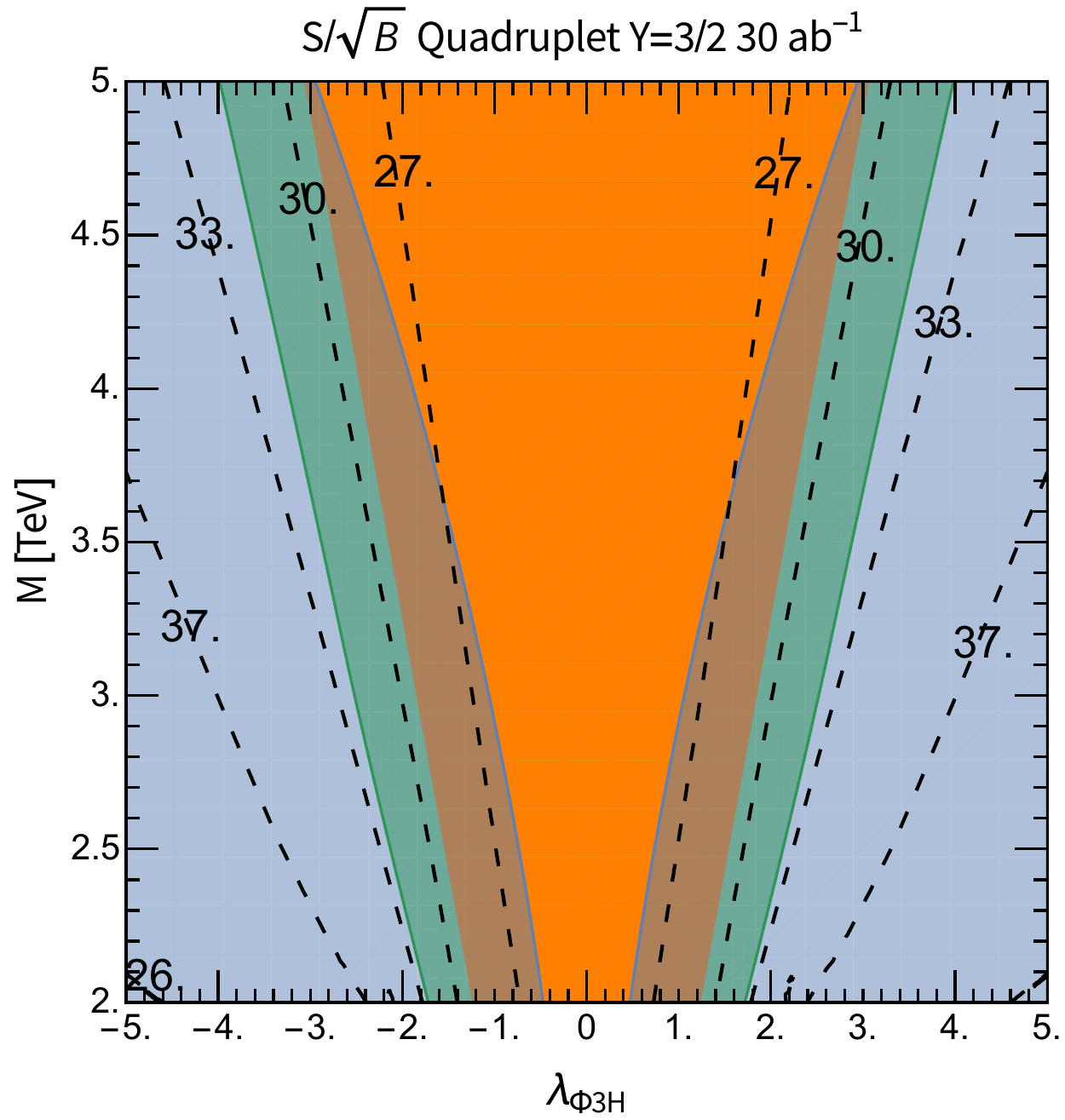}
	\end{center}
	\caption{The discovery potential of the model parameters $(\lambda_{H3\Phi, M})$ in the quadruplet. The dashed black contours correspond to the $S/\sqrt{B}$ values for an integrated luminosity of 30~${\rm ab}^{-1}$. The blue region is excluded by constraints from the electroweak precision tests. The orange and green regions are within the SM  2$\sigma$ uncertainty with an assumption of the percentage theoretical uncertainty of di–Higgs production cross-section equal to 4\% and 10\% respectively.}\label{fig:modelbounds2}
\end{figure}

For the quadruplet model, at dimension-six, only the Wilson coefficient of $Q_H$ operator is non zero. However, we include the $c_{HD}$ generated by dimension 8 operator because it is strongly constraint by EWPD.
In the left plot in Fig.~\ref{fig:modelbounds2}, the allowed region for two types of quadruplet models are denoted by two lines with different slopes.
The reason can be seen from Table.~\ref{tab:EFTtable}, the $c_H$ and $c_{HD}$ are correlated, all proportional to the coupling $|\lambda_{\Phi 3 H}|^2$.
So given a fixed cut off scale $M$, both $g_{HHH}^{(1)}$ and $g_{HHH}^{(2)}$ can be parameterized by a single parameter $|\lambda_{\Phi 3 H}|^2$.
In the right plot in Fig.~\ref{fig:modelbounds2}, we find that the allowed parameter space from the global fit to EWPD for quadruplet models is tightly constrained, and almost becomes a point near the SM value.
In Fig.~\ref{fig:modelbounds}, we show the significance of the model parameter $\lambda_{\Phi 3H}$ vs new physics scale $M$ varies with the $30$ ab$^{-1}$ in contours, while the blue region is excluded by the constraint on $c_{HD}$ from EWPD. One could observe that, the T-parameter constraint on $\lambda_{\phi 3H}$ is very sensitive to the cutoff scale, the reason is that the $c_{HD}$ is generated by dimension-eight operator so that it is proportional to the fourth power of $(v/M)$. 

Our collider analysis demonstrates that the potential of the 100 TeV collider in probing the Wilson coefficients resulting from the five scenarios considered here is very promising with the 2HDM.
The singlet, triplet and quadruplet models on the other hand are restricted due to electroweak precision measurements and their effective coefficients will have less sensitivity. These restrictions also manifest in the constraints on the deviation of the triple Higgs couplings in such models owing to the direct relation between $c_H$ and the triple and quadruplet Higgs coupling as shown in Eq.~\ref{eq:lorentzcoeff}.

An interesting consequence of our analysis is that, 
due to the difference in the allowed region for each model under the theoretical bound and the global fit constraints, it is possible to differentiate the 2HDM model from singlet, triplet and quadruplet models with the observation of a large deviation of the signal rate from the SM expectation. If a future experiment detects a significantly larger signal rate compared with the expected SM model value, then it should favor the presence of an extended scalar sector consisting of the 2HDM the assumptions of this work.
%
%On the other hand, if the future experiment detect a significantly smaller signal rate than the expected SM model value, then it suggests a UV completion of the SM EFT corresponding to the singlet models. We note, however, that both of these statements depend on a lack of new light propagating states which affect the dihiggs signal, an assumption implicit in our use of the EFT framework.
%
If the future experiment does not detect a significant deviation from the SM expectation, then one may have hard time to differentiate SM from all the models considered here as well as models where the wilson coefficients are induced at loop level. Both a reduction in the theoretical uncertainty estimation and higher luminosities will be needed to make a more precise measurement of the di-Higgs signal rate.
%
%Furthermore, depending on the level of deviation one may be able to differentiate between complex singlet model and real singlet model.

%%%%%%%%%%%%%%%%%%%%%%%%%%%%%%%%%%%%%%%%%%%%%%%%%%%%%%%%%%%%%%

%%%%%%%%%%%%%%%%%%%%%%%%%%%%%%%%%%%%%%%%%%%%%%%%%%%%%%%%%%%%%%
\section{Conclusions}
\label{sec:conclusion}
%%%%%%%%%%%%%%%%%%%%%%%%%%%%%%%%%%%%%%%%%%%%%%%%%%%%%%%%%%%%%%

We began by motivating a study of the dimension-six Higgs self-interaction operator  $Q_H=(H^\dagger H)^3$ in the Standard Model effective field theory because of its importance to studying the nature of electroweak symmetry breaking and the nature of the electroweak phase transition. We noted that the largest Wilson coefficients can be obtained by considering extended scalar sectors which are the only models which admit a tree level $Q_H$ operator. After identifying all possible $SU(2)_L$ representations which allow for a tree level $Q_H$ along with the corresponding hypercharge $Y$ we wrote the ultraviolet complete Lagrangians for each model. Finally, assuming that the new scalars are heavy we integrated them out of the theory obtaining the dimension-six effective Lagrangian at tree-level. Of the seven models which generate $Q_H$ at tree level all but two generate more than one effective operators. Those which generate only $Q_H$ at dimension-six are plagued by strong constraints coming from the dimension-eight $T$-parameter operator. This helps put into context that any study performed by shifting a single parameter of the model is not model independent and, in the case of shifts in the self coupling of the Higgs coming from UV physics generating dimension-six operators as leading effects, not well justified.

After identifying the full set of tree-level dimension-six operators for the extended scalar sectors we proceeded to consider the constraints on the effective theories given single Higgs measurements from run I of the LHC as well as the electroweak oblique parameters $S$ and $T$. In order to fully take advantage of the single Higgs measurements we also derived the Wilson coefficient for the effective Higgs coupling to photons although it enters at loop level after integrating out the new heavy charged scalars. For the constraints from the single Higgs measurements we implemented the tool Lilith to perform a global fit to the Wilson coefficients other than the $Q_H$. Having constrained the Wilson coefficients we then projected those constraints back into the  parameters of the ultraviolet models deriving relations between various parameters of the models.

It is the multi-Higgs measurement instead of the single Higgs measurement which determines the Wilson coefficient of the Higgs self-interaction operator $Q_H$. As such, we have investigated the dependence of the coefficient $c_H$ on the di-Higgs production cross section, and studied simulations of the di-Higgs process for the proposed future 100 TeV collider.  We then obtained the sensitivity contours of the Wilson coefficients  in the general effective theory framework as the luminosity varies at the future 100 TeV collider. 
Finally, we reduced the $g_{HHH}^{(1)}-g_{HHH}^{(2)}$ plane, for various cases of the UV complete extended scalar sector models, to its subspace for the cut-off scale of 2 TeV and in the perturbative regime for dimensionless Higgs couplings and demonstrated that most of these regions can be probed to the statistical significance of more than $5\sigma$ using the di-Higgs signatures in a future 100 TeV collider.  

We have converted the discovery reach of the $Q_H$ operator into the Higgs potential parameters in seven UV models. Among the models considered, the Higgs self coupling in the singlet and doublet models could have large deviation from the standard model prediction, while 
the triplet and quadruplet models can only have very small deviation, due to the strong correlation between the $T$ parameter and the Wilson coefficient of the $|H|^6$ operator.
We showed that for the projected data collected for an integrated luminosity of $30$ ab$^{-1}$ at the proposed 100 TeV collider, the trilinear Higgs coupling in the all scalar models with a single heavy scalar integrated out could be fully explored. If a significant deviation in the trilinear Higgs coupling is observed, it will rule out the possibilities of the triplet and quadruplet models. 
On the other hand, if there is only small or no deviation, it will strongly constrain the Higgs potential parameters in the singlet and doublet models.
Therefore, combined with electroweak precision data, the di-Higgs search can effectively differentiate singlet and 2HDM models from triplet and quadruplet models, within the framework of effective field theory of new scalar models.

Overall, the di-Higgs process provides a unique opportunity to probe the $c_H$ operators which cannot be obtained by single Higgs phenomena and the electroweak precision tests.
These future experimental measurements on the Higgs self-interaction operator $Q_H$ will provide important information on the shape of the scalar potential under the assumption that the new scalars are very heavy. This provides a method complementary to direct phenomenological searches to find evidence of additional scalars in these extended scalar models.\\~\\~\\~\\

% {\bf Note added: } As this work was being prepared a similar publication was submitted to the arXiv~\cite{Dawson:2017vgm}. This work contributed to the discussions contained in \cite{Dawson:2017vgm} by expanding on discussions of the 2HDM, including the derivative triple-Higgs coupling corresponding to the coefficient $g_{HHH}^{(2)}$ in Eq.~\ref{eq:lorentzlagrangian} (an approach more suitable for comparison with one-loop analyses and linearly realized gauge symmetry), and a detailed discussion of constraints from EWPD and single Higgs measurements, as well as fully simulated dihiggs production at the proposed future 100 TeV collider. ~\\~\\~\\~\\

\acknowledgments
H.L. and J.H.Y. are supported by the National Science Foundation of China under Grants No. 11875003. T.C. is supported by the Australian Research Council. A.J., H.L.L. and J.H.Y. are supported under U.S. Department of Energy contract DE-SC0011095. J.H.Y. is also supported by the Chinese Academy of Sciences (CAS) Hundred-Talent Program. A.J. is also partially supported by the United States Department of Energy through Grant No. DE FG02 13ER41958.

\appendix

\section{Unitarity considerations}\label{sec:unitarity}

Following the discussion of \cite{Corbett:2014ora,Corbett:2017qgl}, we find unitarity requires that the partial waves for $2\to2$ elastic scattering of bosons be bounded by,
\begin{equation}
|T^J(V_{1\lambda_1}V_{2\lambda_2}\to V_{1\lambda_1}V_{2\lambda_2})|\le2\, ,
\end{equation}
where we may freely substitute $V_{i\lambda_i}\to h$ for the Higgs boson. Considering only amplitudes which grow with the square of the center of mass energy $S$ in the above cited works the authors found that the operator $Q_H$ is not bounded by unitarity considerations for $2\to2$ scattering, and that the operators $Q_{H\Box}$ and $Q_{HD}$ only result in unitarity violation for the purely longitudinal case. Note that as the 2HDM does not generate either $Q_{H\Box}$ or $Q_{HD}$ at leading order in the $Y_3$ expansion it does not generate operators which violate unitarity with growing $S$. It was found that for one operator non-zero at a time the bounds were given by,
\begin{eqnarray}
|c_{H\Box}S|&\le&67\, ,\\
|c_{HD}S|&\le&50\, .
\end{eqnarray}
A simultaneous search of the bounds allowing for cancellations between the two effective couplings yields the bounds:
\begin{eqnarray}
|c_{H\Box}S|&\le&67\, ,\label{eq:simultaneous1}\\
|c_{HD}S|&\le&67\, .\label{eq:simultaneous2}
\end{eqnarray}
It should be noted these constraints indicate the largest allowed values of the two operator coefficients allowing for cancellations between them and not that all values within these bounds will be simultaneously allowed. We may then consider the largest $\sqrt{S}$ at which our EFT is valid (i.e. perturbatively unitary):
\begin{eqnarray}
\sqrt{S}_{\rm crit}&\le&\sqrt{\frac{67}{c_{H\Box}}}\sim\frac{8}{\sqrt{c_{H\Box}}}\, ,\\\
\sqrt{S}_{\rm crit}&\le&\sqrt{\frac{67}{c_{HD}}}\sim\frac{8}{\sqrt{c_{HD}}}\, .
\end{eqnarray}
For di-Higgs processes we consider $\sqrt{S}$ up to 1 TeV, therefore these bounds indicate we should not consider $c_{H\Box}$ or $c_{HD}$ larger than $(8/{\rm TeV})^2$, this bound is well outside the perturbative region of the UV models considered. For example in the case of the real scalar singlet,
\begin{eqnarray}
C_{H\Box}=-\frac{g_{HS}^2}{2M^4}\sim\left(\frac{8}{\rm TeV}\right)^2\, ,
\end{eqnarray}
implies, for $M\sim1$ TeV, that $g\lesssim11$ TeV. For the largest allowed values of $g$ scattering in the UV theory is not perturbative unless $\sqrt{S}>g$, but for $\sqrt{S}> g$ the EFT approach we adopt in this study breaks down.

\bibliography{reference}

\end{document}